\pdfoutput=1
\documentclass{aastex631}

\usepackage{amsthm}
\usepackage{amsmath}
\usepackage{multirow}
\usepackage{hyperref}
\usepackage{float}
\usepackage{subfigure}
\newcommand{\rev}[1]{\color{black}#1 \color{black}}

\begin{document}

\title{Going Beyond the MHD Approximation: Physics-Based Numerical Solution of the CGL Equations}

\correspondingauthor{Deepak Bhoriya}
\email{dbhoriy2@nd.edu}

\author[0000-0003-2849-9045]{Deepak Bhoriya}
\affiliation{Physics Department, University of Notre Dame, Notre Dame, USA}

\author[0000-0003-3309-1052]{Dinshaw Balsara}
\affiliation{Physics Department, University of Notre Dame, Notre Dame, USA}
\affiliation{ACMS Department, University of Notre Dame, Notre Dame, USA}

\author[0000-0001-5485-2872]{Vladimir Florinski}
\affiliation{Department of Space Science, University of Alabama in Huntsville, Huntsville, AL, USA}
\affiliation{Center for Space Plasma and Aeronomic Research, University of Alabama in Huntsville, Huntsville, AL, USA}

\author[0000-0003-4746-2336]{Harish Kumar}
\affiliation{Department of Mathematics, Indian Institute of Technology, Delhi, India}

\begin{abstract}

We present a new numerical model for solving the Chew-Goldberger-Low system of equations describing a bi-Maxwellian plasma in a magnetic field. Heliospheric and geospace environments are often observed to be in an anisotropic state with distinctly different parallel and perpendicular pressure components. The CGL system represents the simplest leading order correction to the common isotropic MHD model that still allows to incorporate the latter's most desirable features. However, the CGL system presents several numerical challenges: the system is not in conservation form, the source terms are stiff, and unlike MHD it is prone to a loss of hyperbolicity if the parallel and perpendicular pressures become too different. The usual cure is to bring the parallel and perpendicular pressures closer to one another; but that has usually been done in an ad hoc manner. We present a physics-informed method of pressure relaxation based on the idea of pitch-angle scattering that keeps the numerical system hyperbolic and naturally leads to zero anisotropy in the limit of very large plasma beta. Numerical codes based on the CGL equations can, therefore, be made to function robustly for any magnetic field strength, including the limit where the magnetic field approaches zero. The capabilities of our new algorithm are demonstrated using several stringent test problems that provide a comparison of the CGL equations in the weakly and strongly collisional limits. This includes a test problem that mimics interaction of a shock with a magnetospheric environment in 2D.

\end{abstract}


\section{Introduction} \label{sec:intro}

\subsection{The MHD system and its limitations}
The magnetohydrodynamic (MHD) equations were first developed by~\cite{alfven1943existence} for describing the large-scale, low-frequency motion of magnetized plasmas. Inbuilt in the MHD single-fluid approximation is the assumption that the plasma is neutral on the scales being considered and that the distributions of ions and electrons that make up the plasma are isotropic, Maxwellian and in equipartition. The assumption of charge neutrality eliminates the Debye frequency which operates on very small scales. The assumption of an isotropic Maxwellian distribution of particle velocities also presumes that the plasma can draw on certain collisional processes to ensure the Maxwellian distribution for the velocities of the electrons and ions on all the scales of interest. The approximation also results in a single density and temperature to describe the thermodynamics as well as the entropy of the particles that make up the plasma. The MHD equations find wide-ranging applicability in astrophysics, space physics, fusion research and many other fields of engineering. Indeed, working within the MHD approximation has served us extremely well! Thanks to the MHD equations, we have models for the magnetized solar wind \citep{weber1967angular, 1975JGR....80.1223S}, the Earth's magnetosphere \citep{zwan1976depletion, pudovkin1977peculiarities}, the heliosphere \citep{1995A&A...304..631B, 1997A&A...321..330P}, the launching of winds and jets \citep{blandford1982hydromagnetic}, dynamo theory \citep{parker2019cosmical, ruzmaikin1988magnetic}, MHD turbulence \citep{sridhar1994toward, goldreich1995toward, goldreich1997magnetohydrodynamic}, the multiphase interstellar medium \citep{mac2005distribution}, and other applications in astrophysics and space physics that are too numerous to mention. Numerical MHD also plays an extremely important role in plasma fusion \citep{robinson2008alegra,hoelzl2021jorek}.

However, it is also well known that the assumption of an isotropic pressure that is inbuilt into MHD is often violated in space plasmas where the Coulomb collision rate is negligible. A compression or expansion of a parcel of plasma associated with a flow can lead to a preferential increase or decrease in the pressure component parallel to the deformation. The best known example is a magnetic field pileup on the day side of the Earth's magnetopause that leads to a condition where the pressure in the directions perpendicular to the field lines exceeds the pressure in the parallel direction -- this is the so-called plasma depletion layer \citep{1979JGR....84..869C, 1993JGR....98.1481G}. Similar depletion layers were subsequently discovered in front of the magnetospheres of other planets \citep{2013JGRA..118.7181G, 2014JGRA..119..121M} and were proposed for the heliopause \citep{2017ApJ...834..197C}. Proton temperatures in the solar wind were also observed to be anisotropic especially during periods an enhanced wave activity \citep{1981JGR....86.9199M, 2001GeoRL..28.2759G, 2006GeoRL..33.9101H}. 

Space plasmas follow a natural trend to maximize their entropy, i.e., relax to an equilibrium state with an isotropic pressure. A collisionless MHD description tacitly assumes strong interaction between particles and magnetic fluctuations causing scattering in pitch angle so frequent, that any departure from isotropy becomes instantly erased. However, in real space environments the mean scattering time is often comparable with the dynamical timescales of the system, and a complete relaxation does not occur. Clearly, standard MHD is not sufficient to model such anisotropic environments numerically. The simplest solution is to assume that the pressure tensor is gyrotropic (axially symmetric about the direction of the magnetic field) and use two pressure variables instead of one; this is the basis of the Chew-Goldberger-Low (CGL) model discussed in the next subsection.

\subsection{CGL as the simplest anisotropic MHD representations} \label{sec:intro_to_plasma}

\cite{1956RSPSA.236..112C} proposed an extension to the MHD system based on conservation of the first two adiabatic invariants of charged particle motion in a magnetic field describing a preservation of magnetic flux in a deformation normal to and parallel to the mean magnetic field, respectively. This approximation is often known as the CGL approximation, and it is also referred to as the double adiabatic approximation. If we take ${D}/{Dt}$ to be the advective derivative, then the CGL theory posits that the pressure perpendicular to the magnetic field, $p_{\perp }$, and the pressure parallel to the magnetic field, ${p}_{\parallel}$, evolve according to
\begin{equation}
\frac{D}{Dt}\left( \frac{{p_{\perp }}}{\rho B} \right)=0,
\quad
\frac{D}{Dt}\left( \frac{{p_{\parallel }} {{B}^{2}}}{{{\rho }^{3}}} \right)=0.
\label{eq:doublead}
\end{equation}
Here $\rho$ is the fluid density and “$B$” is the magnitude of the magnetic field. The CGL equations govern large (fluid) scale phenomena, where the physical processes take place on timescales that are much larger than the proton gyro-period \rev{ and varies over distances much greater than the Larmor radius \citep[e.g.,][]{1969fecg.book.....B}.}
	
To illustrate the effects of the above two equations, as they relate to the double adiabatic approximation, let us consider a simple thought experiment. Let us consider a flux tube with length “L” and area “$\Sigma $”. Then, as the flux tube is stretched or compressed, we have $B\propto {{\Sigma }^{-1}}$and $\rho \propto {{\left( L\Sigma  \right)}^{-1}}$. With these scalings, we can also define the temperature perpendicular and parallel to the flux tube as ${{T}_{\parallel }}\equiv {{{p}_{\parallel }}}/{\rho }\;$ and ${{T}_{\perp }}\equiv {{{p}_{\perp }}}/{\rho }\;$. We then see from the above two equations that ${{T}_{\parallel }}\propto {{L}^{-2}}$ and ${{T}_{\bot }}\propto {{\Sigma }^{-1}}$, showing that, as the flux tube is stretched or compressed, different types of changes are brought about in the parallel and perpendicular temperatures. The CGL formalism neglects kinetic effects leading to collective interaction between particles and waves, and therefore admits unrestricted values of the anisotropy in pressures, defined as $A\equiv {{{p}_{\perp }}}/{{p_{\parallel }}}\;$.

In reality, particles are affected by electromagnetic fluctuations, either of ambient origin (what is commonly known as turbulence) or self-generated as a result of an instability \citep{1958PhRv..109.1874P}. Several types of plasma instabilities can arise in a plasma with an anisotropic pressure distribution \citep{chandrasekhar1958stability, barnes1966collisionless, kennel1966limit}. These processes place constraints on the range of anisotropy values attainable in space plasmas. Thus, the physics of producing anisotropy (via the double adiabatic approximation) as well as the physics of restoring isotropy (via inclusion of the physics of the above-mentioned instabilities and turbulence) needs to be put into our large-scale models for astrophysical and space plasmas. The goal of this paper is to deliver on this need.

Formally, the only difference between the MHD and CGL system of equations is the representation of the pressure tensor. Whereas for the former $\mathbf{p}=p\mathbf{I}$, where $\mathbf{I}$ is the identity matrix, the latter has a diagonal form in the coordinate frame aligned with the mean magnetic field. While that form is convenient for particle-based models, it is not useful in most grid-based models that rely on a fixed coordinate frame. In that frame, the expression for the pressure tensor becomes
\begin{equation}
\label{eq:ptensor}
\mathbf{p}=p_\perp\mathbf{I}+(p_\parallel-p_\perp)\mathbf{bb},
\end{equation}
where $\mathbf{b}=\mathbf{B}/B$ is the unit vector in the direction of the magnetic field. To describe the relative strength of thermal effects compared with magnetization effects in the double adiabatic approximation, two plasma beta parameters and their average are introduced,
\begin{equation}
\label{eq:beta}
\beta_\parallel=\frac{8\pi p_\parallel}{B^2}, \quad\beta_\perp=\frac{8\pi p_\perp}{B^2}, \quad\bar{\beta}=\frac{\beta_\parallel+2\beta_\perp}{3}.
\end{equation}
Note that expression (\ref{eq:ptensor}) is vaguely defined in the limit of a very weak magnetic field, which has in the past led to numerical difficulties in CGL codes. We will show in this paper (subsection 4.3) that a physics-based approach overcomes this difficulty.

\subsection{Previous work and the guidance it provides}

The CGL system consists of five equations, three of which are scalar and two are vector equations. Most authors use equations for the density, velocity, and magnetic field in a conservative form (similar to MHD). However, there is no accepted standard for incorporating equations (\ref{eq:doublead}) into the system, and multiple approaches can be found in the literature. The paper of \citet{1992JGR....9710643H} represents an early implementation of the CGL system in a three-dimensional numerical model of magnetotail dynamics. The authors used evolution equations for $p_\parallel$ and $p_\perp$, while the total energy was not explicitly conserved. It was also recognized that a relaxation term was needed to keep the anisotropies from becoming unphysically large. \citet{1992JGR....9710643H} in particular used a relaxation term that drove the system toward a marginally stable state on a timescale $\tau$ given by the linear growth rates of the temperature anisotropy plasma instabilities (the firehose and mirror instabilities, see below). \citet{1996GeoRL..23.2891D} developed a 2D model CGL model for magnetosheath simulations that evolved the adiabatic invariants (\ref{eq:doublead}) directly. The paper also introduced a bounded anisotropy model, where instead of a relaxation term the code re-adjusted $p_\parallel$ and $p_\perp$ to the nearest state of marginal stability (MS) without affecting the trace of the pressure tensor at the end of each time step. An improved version was presented in \citet{2000JGR...105.7545D}.

A steady state 3D model of the day side magnetosheath was published by \citet{1999JGR...104.6877E}. These authors also used the instability threshold conditions as a ``fence'' preventing the system from leaving the stability region. \citet{1999PhPl....6.2887P} investigated pressure anisotropies in the magnetosheath with a steady state 2D code. A different technique was demonstrated for anisotropy reduction where the relaxation time was based on pitch angle scattering time. In this approach, the pressure ratio approaches unity with time instead of remaining in the state of marginal stability. To avoid the difficulties in calculating the scattering rates from physical principles the model used a constant scattering time as a free parameter. \citet{2001JGR...10621689S} argued for the existence of a magnetopause plasma depletion layer based on the results from their 3D time-dependent anisotropic MHD model. That model solved the energy conservation law plus a separate equation for the anisotropy $A$. A threshold method of isotropization was used with $\tau$ again treated as a free parameter. Note that the papers discussed so far were specifically intended for magnetospheric simulations rather than for general use.

The first general framework for the CGL-MHD system was presented by \citet{2012JCoPh.231.3610M}. The model evolved parallel and perpendicular pressure equations in a non-conservative form. The properties of the characteristic matrix of the system of 9 equations and the corresponding eigenvalues (wave speeds) were discussed in detail. The relaxation method was also based on instabilities, and the system was driven toward the nearest stable state. The relaxation time was taken to be a constant for simplicity. The next revision of this model \citep{2012JGRA..117.8216M} used an improved relaxation time based on the growth rate of the most unstable modes. \citet{2014ApJ...781...84S} studied turbulence on very large (cosmological) scales with a model that combined eqns. (\ref{eq:doublead}) into a single conservation law for the quantity $A(\rho/B)^3$. Note that this quantity can become very large in the gas-dynamic limit ($\beta\to\infty$), which would be undesirable for general-purpose models. Pressure relaxation was applied separately from the main scheme by transforming to primitive variables; the relaxation time was a free parameter. The authors pointed out that non-Hall MHD models do not resolve gyroscales, and therefore, an attempt to incorporate a kinetic wave growth scale would be very similar to a ``hard fence'' approach. The updated model \citep{2016MNRAS.460.2492S} incorporated anisotropy relaxation terms derived from quasi-linear theory \citep{2012JGRA..117.8101S, 2012JGRA..117.8102Y}. It was found that the rate of anisotropy creation via a flow deformation is, in most cases, an extremely small fraction of the scattering rate, a fact lending support to the simple bounded anisotropy model.

\rev{The above-mentioned papers used a Riemann solver to compute the interface fluxes in their numerical codes. An alternative approach, presented in \citet{2023JCoPh.49012311L}, developed a kinetic flux-splitting method of flux calculation by taking moments of a bi-Maxwellian distribution. The model was tested on the magnetic reconnection problem driven by numerical dissipation. \citet{Sharma_Hammett_Quataert_Stone_2006} extended the double adiabatic conservation laws with an addition of the heat flux. In expressing the heat flux components through lower order moments they use a Landau fluid closure, but characterized by a single wavelength that was a multiple of the grid cell size. Unlike conventional CGL, that mode correctly predicts the growth rate of the mirror instability. The authors used more elaborate relaxation times that incorporated the instability growth rates on the scales comparable with particle's Larmor radius, but the resulting relaxation method was conceptually similar to the hard fence. \citet{Hu_Denton_Lin_2010} compared the solutions to the CGL + Landau fluid heat flux system to a kinetic particle-mesh simulation. They reported substantial differences between the parallel heat fluxes in the two models; but otherwise their results suggested that the fluid model in general is a viable alternative to the kinetic approach for the Earths's magnetosphere.}

A different approach to anisotropic MHD that does not involve solving the CGL system was proposed by \cite{2016JCoPh.327..851H}. Their model evolved the ten-moment one fluid set of equations incorporating a general non-gyrotropic pressure tensor. Gyrotropization was achieved by a relaxation term. While this method has the advantage of being able to readily deal with situations where $B\to 0$, it is more cumbersome to implement and does not preserve the familiar characteristics of the MHD system of equations. The detailed eigenstructure for this system is also not available making it difficult to design very stable numerical schemes that rely on reconstruction of the eigenweights.

The present paper seeks to develop a general-purpose formalism for the CGL system of equations that can be applied to a broad range of anisotropic space plasma environments with (in principle) arbitrary values of the plasma beta, including the gas-dynamic limit. The model strictly enforces the conservation of energy and preserves the divergence-free nature of the magnetic field, which was not addressed in earlier developments. In addition, our model introduces an ``elastic fence'' approach to anisotropy relaxation wherein the system tends to equilibrate the pressures at a rate that is driven by a physical relaxation time for most values of the anisotropy, as long as the anisotropy is in the physical range (i.e. inside the fence). However, for situations where the system comes close to the hard boundaries of the fence, it is smoothly allowed to relax back at a faster rate so that the system never fully reaches the ``hard fence''. There is a logical reason for this compromise: If the anisotropy reaches the hard fence, the hyperbolic system loses its physical realizability -- the eigenvalues of the system change character from real to imaginary! It is, therefore, safer to keep a small margin which prevents the anisotropy from reaching the hard fence. If the system approaches the hard fence, our strategy is to modify the relaxation time to smoothly move the system away from the hard fence. This ensures that the physical variables in the hyperbolic system always stay within the domain of hyperbolicity. In later sections we describe how special care is taken in the implementation of this ``elastic fence'' to ensure that the mathematical system remains within physical bounds. Interestingly, we show that this also provides a salutary cure to the CGL equations when the magnetic field tends to zero; i.e., our strategy smoothly retrieves an isotropic pressure for the CGL equations in the limit when the magnetic field tends to zero.

\section{The CGL Equations and the Numerical Challenges that they Pose} \label{sec:systemcgl}
The CGL equations can be written in the cgs units as a non-conservative hyperbolic system with a stiff source term as follows:
\begin{align} 
\frac{ \partial \rho}{\partial t}+\nabla \cdot (\rho \mathbf{v})&=0
\label{eq:mass_conv} \\
\frac{\partial (\rho \mathbf{v})}{\partial t}
+
\nabla \cdot 
\left[\rho \mathbf{v} \mathbf{v}
+p_\perp \textbf{I}+(p_\parallel-p_\perp) \mathbf{ b} \mathbf{ b} -\frac{1}{4\pi}
\left(\mathbf{B}\mathbf{B}-\frac{B^2}{2}\textbf{I}\right) \right]
&=
0
\label{eq:momentum_conv} \\
\frac{\partial (p_\parallel-p_\perp)}{\partial t}
+
\nabla \cdot \left[(p_\parallel-p_\perp)\mathbf{v}\right] +(2p_\parallel+p_\perp )\mathbf{ b}\cdot\nabla \mathbf{v}\cdot\mathbf{ b} - p_\perp\nabla\cdot\mathbf{v}
&=
-\frac{1}{\tau} \left( p_\parallel - p_\perp \right)
\label{eq:press_jump} \\
\frac{\partial \mathcal{E}}{\partial t} +
\nabla \cdot \left[\mathbf{v}\left(\mathcal{E} + p_\perp+ \frac{B^2}{8 \pi}\right)
+
\mathbf{v} \cdot \left((p_\parallel-p_\perp) \mathbf{ b} \mathbf{ b} -\frac{\mathbf{B} \mathbf{B}}{4 \pi}\right)\right]&=0
\label{eq:energy_conv}\\
\frac{\partial \mathbf{B}}{\partial t}+\nabla \times \left[-(\mathbf{v} \times \mathbf{B})\right]&=0
\label{eq:induction}
\end{align}
In the above set of equations, $\rho$ represents the total density and $\mathbf{v} = (v_x, v_y, v_z)^\top$ denotes the velocity vector. The relaxation time is denoted by $\tau$. The system is closed under the following equation of state
\begin{equation}
    \mathcal{E}=\dfrac{1}{2}\rho {v}^2 + \frac{B^2}{8 \pi} + \frac{3}{2} \bar{p}
    \label{eq:EOS}
\end{equation}
where $\bar{p}\equiv \dfrac{2 p_\perp + p_\parallel}{3}$ is the average pressure. \rev{Our approach of solving for the pressure difference is uncommon in the literature, but we never experienced any difficulties with this formulation as can be seen from the results presented in \S\ref{sec:prob}.}

We see that eqns.~\eqref{eq:mass_conv} through~\eqref{eq:energy_conv} evolve zone-centered variables; however, eqn.~\eqref{eq:induction} for the magnetic field requires the evolution of facial magnetic fields. Since we want a general formulation for the CGL equations that can hold at all orders of accuracy, we will have to pay attention to the non-oscillatory reconstruction of zone-centered variables as well as face-centered variables. The zone-centered variables can be reconstructed via well-known Weighted Essentially Non-Oscillatory (WENO) schemes. For a very efficient implementation of these schemes, please see the supplement in~\cite{balsara2023efficient}. The reconstruction of face-centered variables has been described in~\cite{balsara2001divergence,balsara2004second,balsara2009divergence} and~\cite{balsara2018computational}. Therefore, while the implementation of such methods may be slightly challenging, we will regard it as a solved problem.

Once the solution has been reconstructed in space, it has to be evolved in time. To keep our approach as simple as possible, we will rely on multi-stage Runge-Kutta methods; but the next paragraph will show that the CGL system has intricacies which require a special form of Runge-Kutta timestepping method. Realize though that every stage in any Runge-Kutta method will, in the very least, require the application of a one-dimensional Riemann solver at zone boundaries. We see that the mass, momentum and total energy equations, i.e. eqns.~\eqref{eq:mass_conv},~\eqref{eq:momentum_conv} and~\eqref{eq:energy_conv} respectively, are in flux conservation form. Such PDEs that are in flux conservation form are familiar to the astrophysics community. However, notice that eqn.~\eqref{eq:press_jump} for the evolution of the difference between the parallel and perpendicular pressures, is not in conservation form. Riemann solvers that can handle non-conservative products in the hyperbolic PDE are not so well-known in the astrophysics community. We will, therefore, need a Riemann solver that is designed to accommodate non-conservative products \citep{dumbser2016new}. In such circumstances, the Riemann solver has to return fluctuations and the PDE has to be updated in fluctuation form. The presence of a non-conservative product implies that the Riemann solver becomes more complicated, and more expensive because it has to be iterated to convergence. Fortunately, much of the PDE system is indeed in conservation form, so our method should be such as to preserve conservation whenever it is mandated by the governing equations. Even so, our method should be flexible enough to provide proper upwinded treatment for the non-conservative products. That is one of the challenges that we will have to face in this paper when formulating methods for the CGL equations. A known solution has been presented in Appendix C of~\cite{dumbser2016new} in the form of the HLLEMNC pseudocode. As a result, while this may be challenging to implement, it is viewed here as a challenge that can be met, and has a good resolution in the literature. 

Notice too that the induction equation, eqn.~\eqref{eq:induction}, requires that facial magnetic fields be updated with edge-centered electric fields. The electric field will have to be obtained from a multidimensional Riemann solver. Multidimensional Riemann solvers will therefore be needed for obtaining the electric field at the edges of the mesh. Fortunately, the multidimensional Riemann problem for CGL is not any different from the multidimensional Riemann problem for MHD \citep{balsara2010multidimensional, balsara2012two, balsara2014multidimensional, balsara2017multidimensional}; therefore, this is regarded as a challenge that is resolved.

We also see that the right-hand side of eqn.~\eqref{eq:press_jump} has a relaxation term. When the relaxation timescale ``$\tau$" (tau) becomes small, the parallel and perpendicular pressures must approach one another. Note however, that ``$\tau$" will not be a pre-specified constant; instead it will depend on the physics of the problem and will change as the local flow variables change in each zone. Since the relaxation timescale can become solution-dependent, one cannot use simpler exponential time differencing-based Runge-Kutta methods. Observe from the right-hand side of eqn.~\eqref{eq:press_jump} that when the difference in pressures is non-zero and the relaxation time is small, a stiffly stable method for treating the source terms is not guaranteed to preserve the sign of the pressure difference! Allowing the source terms to change the sign of the pressure difference can cause unphysical oscillations in the pressure. Furthermore, one class of plasma instabilities operates when the parallel pressure is greater than the perpendicular pressure and another class of plasma instabilities operates when the parallel pressure is less than the perpendicular pressure. This shows us that it is important for the source term treatment to preserve the sign of $\left( p_\parallel - p_\perp \right)$. We, therefore, need a special type of Runge-Kutta treatment of source terms which guarantees that the sign of the pressure difference cannot be changed by the source terms acting by themselves. (Note though that the left-hand side of eqn.~\eqref{eq:press_jump} can indeed change the sign of the pressure difference.) In other words, we need a sign-preserving Runge-Kutta treatment of source terms. That is another challenge associated with the CGL equations. While such sign-preserving Runge-Kutta schemes have appeared in the recent literature \citep{huang2018bound}, this idea has not percolated into the astrophysical literature, and we will provide some detail here.

It should also be observed that the flux terms and other spatial gradient terms in eqns.~\eqref{eq:mass_conv},~\eqref{eq:momentum_conv} and~\eqref{eq:press_jump} require an evaluation of the unit vector, “$\mathbf{b}\equiv {\mathbf{B}}/B$”, that is in the direction of the magnetic field. This requires that a well-defined magnetic field should be present at all points within a zone. As the magnetic field tends to zero, parts of those flux terms could become singular. Therefore, an approach has to be found to prevent that.

Most importantly, the CGL system is prone to a loss of hyperbolicity. The loss of hyperbolicity arises when the pressure anisotropy $A\equiv {{{p}_{\perp }}}/{{{p}_{\parallel }}}\;$ becomes much larger than unity or much smaller than unity. As \rev{shown in~\cite{kato1966propagation}}, the extent or the anisotropy that can be tolerated without loss of hyperbolicity is determined by the strength of the local magnetic field. We examine this issue further in \S \ref{sec:eigenvalue}.

\section{Innovations in Numerical Methods for the CGL Equations} \label{sec:numcgl}
We split this section into three sub-sections. The first subsection shows how the timestepping is modified to yield a sign-preserving update strategy that can accommodate the hyperbolic terms while treating the source term in a fashion that is sign-preserving and unconditionally stable. The second subsection shows how this is integrated into a scheme that is written in fluctuation form; which can, therefore, accommodate non-conservative products. The third subsection catalogues the major results from an eigenvalue analysis of the CGL system which also reveals the limits where the system loses hyperbolicity.

\begin{figure}[h]
\begin{center}
\includegraphics[width=0.5\textwidth,clip=]{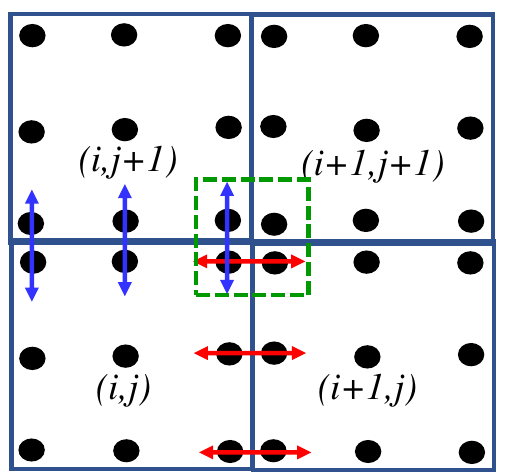}
\caption{The arrangement of the spatial nodes in the third or fourth order accurate RK-WENO algorithm for two scaled space dimensions. The nodes within four abutting spatial zones are shown by the black dots. The red, double-sided arrows indicate the application of  1D Riemann solvers at the nodal points in the x-direction. The blue, double-sided arrows indicate the application of  1D Riemann solvers at the nodal points in the y-direction. The green dashed square at the right-upper vertex of zone $(i,j)$ indicates the application of 2D Riemann solvers at the vertices of the mesh.}
\label{fig:grid}
\end{center}
\end{figure}

\subsection{The PDE System and Sign-Preserving, Stiffly Stable, Runge-Kutta Time stepping} \label{sec:PDE_time_stepping}
To keep the description tractable, we show the steps in two spatial dimensions. Once the two-dimensional formulation is understood, the three-dimensional extension is obvious. In 2D, we assume that we have a hyperbolic PDE with non-conservative products of the form
\begin{align}
   {{\partial }_{t}}\mathbf{U}+{{\partial }_{x}}\mathbf{F}\left( \mathbf{U} \right)+{{\partial }_{y}}\mathbf{G}\left( \mathbf{U} \right)+\mathbf{P}\left( \mathbf{U} \right){{\partial }_{x}}\mathbf{U}+\mathbf{Q}\left( \mathbf{U} \right){{\partial }_{y}}\mathbf{U}=\dfrac{1}{\tau }\mathbf{S}\left( \mathbf{U} \right)
\label{eq:PDE}
\end{align}
Here “$\mathbf{U}$” is the solution vector, “$\mathbf{F}\left( \mathbf{U} \right)$” and “$\mathbf{G}\left( \mathbf{U} \right)$” are the x- and y-fluxes, “$\mathbf{P}\left( \mathbf{U} \right)$” and “$\mathbf{Q}\left( \mathbf{U} \right)$” are matrices with non-conservative products and “$\mathbf{S}\left( \mathbf{U} \right)$” is a source term, assumed to be stiff. It is easy to see that Eqns.~\eqref{eq:mass_conv} to \eqref{eq:induction} have this form where the fluxes in the x- and y-directions are easily identified and the non-conservative products in Eqn.~\eqref{eq:press_jump} can be written very simply as matrices “$\mathbf{P}\left( \mathbf{U} \right)$” and “$\mathbf{Q}\left( \mathbf{U} \right)$”. We emphasize that the source is stiff by identifying ${\mathbf{S}\left( \mathbf{U} \right)}/{\tau }\;$ in Eqn.~\eqref{eq:PDE} with the term $-{\left( {{p}_{\parallel }}-{{p}_{\bot }} \right)}/{\tau }\;$ in Eqn.~\eqref{eq:press_jump}. Notice that in order to introduce a physics-based “elastic fence”, we will have to eventually make “$\tau $” solution-dependent. To write the Runge-Kutta scheme compactly, we have to compress the notation so that we write the update in Eqn.~\eqref{eq:PDE} as
\begin{align}
{\partial _t}{\mathbf{U}} = \mathcal{L}\left( {\mathbf{U}} \right) + \frac{1}{\tau }{\mathbf{S}}\left( {\mathbf{U}} \right)
  \text{\qquad with \qquad}
\mathcal{L} \left( {\mathbf{U}} \right) \equiv  - {\partial _x}{\mathbf{F}}\left( {\mathbf{U}} \right) - {\partial _y}{\mathbf{G}}\left( {\mathbf{U}} \right) - {\mathbf{P}}\left( {\mathbf{U}} \right){\partial _x}{\mathbf{U}} - {\mathbf{Q}}\left( {\mathbf{U}} \right){\partial _y}{\mathbf{U}}
\label{eq:operator_L}
\end{align}

In order to have an “elastic fence” that keeps \(\left( {{p_\parallel } - {p_ \bot }} \right)\) within the physically allowed bounds, we will need a method that can treat the source term in an unconditionally stable fashion, even in the limit where the relaxation time \(\tau  \to 0\). This property is called stiff stability. While several approaches for treating the source term can be stiffly stable, many of them will not preserve the sign of \(\left( {{p_\parallel } - {p_\bot }} \right)\) when the source term, and only the source term, is applied to the PDE system. If the source term integrator is not sign-preserving, the sign of \(\left( {{p_\parallel } - {p_\bot }} \right)\) can change even when the source is treated by itself. The unhappy consequence of that is that the source term treatment can, by itself, cause oscillations in \(\left( {{p_\parallel } - {p_\bot }}\right)\), which runs counter to the physics of the CGL equations. We want a stiffly stable treatment of the source terms that is sign-preserving. Technically, such schemes are known as bound-preserving schemes. Such bound-preserving schemes were designed with second and third-order temporal accuracy by \cite{huang2018bound}. The trick is to make the timestepping as close as possible to an exponential time differencing scheme, which is guaranteed to be bounds preserving because of the very nature of the exponential function. (In the next section we also describe how the relaxation time “\(\tau \)” is made solution-dependent in order to construct a “soft fence” that is physics-based.) We make a small modification of \cite{huang2018bound} to obtain a bounds-preserving, two-stage, second-order, Strong Stability Preserving Runge-Kutta scheme that takes the solution from a state vector \({{\mathbf{U}}^n}\) at time \({t^n}\) to a state \({{\mathbf{U}}^{n + 1}}\) at time \({t^{n + 1}} = {t^n} + \Delta t\), can be written as:
%
\setlength{\abovedisplayskip}{10pt}
\setlength{\belowdisplayskip}{10pt}
\begin{subequations}
\begin{align}
    {\mathbf{U}}^{(1)} &= \dfrac{1}{1+z + \dfrac{1}{2}z^2}
    \left(
    {\mathbf{U}}^n + \Delta t \mathcal{L} (\mathbf{U}^n) 
    \right),
    \\
    {\mathbf{U}}^{n+1} &= \dfrac{1}{ 2 \left( 1+z + \dfrac{1}{2}z^2 \right) }
     \textbf{U}^n
     +
     \dfrac{1}{2}
    \left(
    {\mathbf{U}}^{(1)} + \Delta t \mathcal{L} (\mathbf{U}^{(1)}) 
    \right),
\end{align}
where $z=\dfrac{\Delta t}{\tau}$.
    \label{eq:timeo2}
\end{subequations}
We are assuming that we have a linear source term within each zone; however the relaxation time can vary from one zone to the next. Notice that the scheme is explicit in the treatment of the source term, a favorable property that it derives from the original exponential time differentiation. The corresponding three-stage, third-order, Strong Stability Preserving Runge-Kutta scheme can be written as:-
\begin{subequations}
\label{eq:timeo3}
\begin{align}
    {\mathbf{U}}^{(1)} &= \dfrac{1}
    {1+z+\dfrac{1}{2}z^2 + \dfrac{1}{6} z^3}
    \left(
    {\mathbf{U}}^n + \Delta t \mathcal{L} (\mathbf{U}^n)
    \right),
    \\
    {\mathbf{U}}^{(2)} &= \dfrac{3}{ 4 \left( 
    1+\dfrac{1}{2}z+\dfrac{1}{8}z^2 + \dfrac{1}{48} z^3 \right)}
     \textbf{U}^n 
     +
     \dfrac{1+z+\dfrac{1}{2}z^2 + \dfrac{1}{6} z^3}{4 \left(
     1+\dfrac{1}{2}z+\dfrac{1}{8}z^2 + \dfrac{1}{48} z^3
     \right)}
    \left(
    {\mathbf{U}}^{(1)} + \Delta t \mathcal{L} (\mathbf{U}^{(1)})
    \right),
    \\
    {\mathbf{U}}^{n+1} &= \dfrac{1}{ 3 \left( 
    1+z+\dfrac{1}{2}z^2 + \dfrac{1}{6} z^3 \right)}
     \textbf{U}^n 
     +
     \dfrac{2\left(1+\dfrac{1}{2}z+\dfrac{1}{8}z^2 + \dfrac{1}{48} z^3 \right) }{3 \left(
     1+z+\dfrac{1}{2}z^2 + \dfrac{1}{6} z^3
     \right)}
    \left(
    {\mathbf{U}}^{(2)} + \Delta t \mathcal{L} (\mathbf{U}^{(2)})
    \right),
\end{align}
where $z\equiv\dfrac{\Delta t}{\tau}$.
\end{subequations}
%
%
%
%
\subsection{Detailed Description of a Single Stage in the Update} \label{sec:single update}
We realize that a higher-order WENO reconstruction can be constructed for the zone-centered variables as well as the face-centered variables using the WENO methods that are cited soon after eqn.~\eqref{eq:EOS}. This gives us a spatially high-order representation of the solution within each zone of the computational domain. Now please see Figure~\ref{fig:grid}, which shows four contiguous zones along with the fourth-order accurate nodal points that are used in each zone. The nodal points can be integrated in any one dimension using the Simpson rule and in multiple dimensions using a tensor product of the Simpson rule. Within each physical zone, we map \(\left( {x,y} \right) \in \left[ { - \Delta x/2,\Delta x/2} \right] \times \left[ { - \Delta y/2,\Delta y/2} \right]\) to the reference element \(\left( {\xi ,\psi } \right) \in \left[ { - 1/2,1/2} \right] \times \left[ { - 1/2,1/2} \right]\) so that all points within a zone can be accessed by ranging from $-1/2$ to $1/2$ in each of the local coordinate directions for that zone. In eqns.~\eqref{eq:timeo2} and~\eqref{eq:timeo3} we have already described the temporal update. Our task in this section is, therefore, to describe the spatial update by describing how \(\mathcal{L}\left( {\mathbf{U}} \right)\) is obtained in any zone.

Now imagine four zones like the ones in Figure~\ref{fig:grid} being placed side by side. We realize that at each interface between any two neighboring zones we have two sets of states, one from one side of the zone boundary and the other from the other side of the zone boundary. These two states can be sent as input states into a one-dimensional Riemann solver. The Riemann solver gives us information about the wave contributions that move to the left of the zone boundary and to the right of the zone boundary. Appendix C of \cite{dumbser2016new} gives us one example of such a Riemann solver that can work with PDEs in conservation form as well as PDEs with non-conservative products. The fluctuation formulation then gives us an equivalent view of how to extract the wave contributions that move to the left of the boundary and to the right of the boundary. The Riemann solver applied in the x-direction at a zone boundary gives us two sets of fluctuations. At each x-face of the mesh we get left-going x-directional fluctuations at the nodal points in that face, which we denote as \({\mathbf{D}^{ - x}}\), as well as right-going fluctuations in the same face, which we denote as \({\mathbf{D}^{ + x}}\). These fluctuations in either x-direction are shown by the red double-sided arrows in Figure~\ref{fig:grid}. Likewise, at each y-face of the mesh we get downward-going y-directional fluctuations at the nodal points in that face, which we denote as \({\mathbf{D}^{ - y}}\), as well as upward-going y-directional fluctuations in the same face, which we denote as \({\mathbf{D}^{ + y}}\). These fluctuations in either y-direction are shown by the blue double-sided arrows in Figure~\ref{fig:grid}. We can now write the update for the zone-centered variables as
\begin{equation}
\begin{aligned}
    \mathcal{L} \left( {\mathbf{U}} \right) =&  - \frac{1}{{\Delta x}}\left\{ {\int\limits_{ - 1/2}^{1/2} {\left[ {{\mathbf{D}}_{}^{ - x}\left( {\xi  = 1/2,\psi } \right)} \right]d\psi }  + \int\limits_{ - 1/2}^{1/2} {\left[ {{\mathbf{D}}_{}^{ + x}\left( {\xi  =  - 1/2,\psi } \right)} \right]d\psi } } \right\}
    \\
    & -  \frac{1}{{\Delta y}}\left\{ {\int\limits_{ - 1/2}^{1/2} {\left[ {{\mathbf{D}}_{}^{ - y}\left( {\xi ,\psi  = 1/2} \right)} \right]d\xi }  + \int\limits_{ - 1/2}^{1/2} {\left[ {{\mathbf{D}}_{}^{ + y}\left( {\xi ,\psi  =  - 1/2} \right)} \right]d\psi } } \right\}
    \\
    & - \frac{1}{{\Delta x}}\left\{ {\int\limits_{ - 1/2}^{1/2} {\left[ {{\mathbf{F}}\left( {\xi  = {{\left( {1/2} \right)}^ - },\psi } \right)} \right]d\psi }  - \int\limits_{ - 1/2}^{1/2} {\left[ {{\mathbf{F}}\left( {\xi  =  - {{\left( {1/2} \right)}^ + },\psi } \right)} \right]d\psi } } \right\}
    \\
    & - \frac{1}{{\Delta y}}\left\{ {\int\limits_{ - 1/2}^{1/2} {\left[ {{\mathbf{G}}\left( {\xi ,\psi  = {{\left( {1/2} \right)}^ - }} \right)} \right]d\xi }  - \int\limits_{ - 1/2}^{1/2} {\left[ {{\mathbf{G}}\left( {\xi ,\psi  =  - {{\left( {1/2} \right)}^ + }} \right)} \right]d\psi } } \right\}
    \\
    & - \frac{1}{{\Delta x}}\int\limits_{ - 1/2}^{1/2} {\int\limits_{ - 1/2}^{1/2} {{\mathbf{P}}\left( {\xi ,\psi } \right){\partial _\xi }{\mathbf{U}}\left( {\xi ,\psi } \right)d\xi } d\psi }  - \frac{1}{{\Delta y}}\int\limits_{ - 1/2}^{1/2} {\int\limits_{ - 1/2}^{1/2} {{\mathbf{Q}}\left( {\xi ,\psi } \right){\partial _\psi }{\mathbf{U}}\left( {\xi ,\psi } \right)d\xi } d\psi }
\end{aligned}
\label{eq:Lupdate}
\end{equation}
Also notice from the right upper vertex of zone \(\left( {i,j} \right)\) in Figure~\ref{fig:grid} that we have four states surrounding that vertex. As a result, those four states can be sent into a multidimensional Riemann solver \cite{balsara2010multidimensional,balsara2012two,balsara2014multidimensional,balsara2017multidimensional} and they produce a multidimensionally upwinded representation of the electric field at the vertex. An update for the facial magnetic fields can, therefore, be similarly designed and this is well-described in the above-mentioned references as well as \cite{balsara1999staggered}. Having described all the details for a single stage in the Runge-Kutta update in this subsection, as well as having described how the different stages have to be coupled together in the previous subsection, we can now conclude that the full spatially and temporally high-order scheme has been described here. \rev{In this Sub-section we have described the method at a conceptual level. In Sub-section~\ref{sec:implementation} we provide a step-by-step implementation plan.}
\subsection{Eigenvalue analysis for the CGL equations}
\label{sec:eigenvalue}
The set of eigenvalues plays a crucial role in the design of efficient numerical schemes. For that reason, we provide the set of eigenvalues (denoted by $\Lambda_d$, $d=x,y,z$) for the $d-$directional characteristic matrix below.
\begin{equation}
    \Lambda_d = \{v_d - m_f, v_d - m_s, v_d - c_a, v_d, v_d, 0, v_d + c_a, v_d + m_s, v_d + m_f\},
\end{equation}
where
\begin{align*}
    c_a &= \sqrt{\frac{{B_d }^2 }{4\,\rho \,\pi }-\frac{{b_d }^2 \,{\left(p_{\parallel} -p_{\perp} \right)}}{\rho }},
    \ (\text{where} \ b_d = B_d / |\mathbf{B}|)
    \\
    m_s &= \sqrt{\frac{-\mathrm{b}-\sqrt{{\mathrm{b}}^2 -4\,\mathrm{a}\,\mathrm{c}}}{2\,\rho }}, \\
    m_f &= \sqrt{\frac{-\mathrm{b}+\sqrt{{\mathrm{b}}^2 -4\,\mathrm{a}\,\mathrm{c}}}{2\,\rho }}.
\end{align*}
The expressions for the unknowns `a', `b' and `c' in the above equations are given by
\begin{align*}
    \mathrm{a} &= 2\rho
    \\
    \mathrm{b} &= -2\,p_{\perp} -{b_d }^2 \,{\left(2\,p_{\parallel} -p_{\perp} \right)}-\frac{B^2}{4\,\pi }
    \\
    \mathrm{c} &= -\frac{3\,{b_d }^4 \,{p_{\parallel} }^2 -{b_d }^2 \,{p_{\perp} }^2 \,
          {\left({b_d }^2 -1\right)}
          -\dfrac{3\,{B_d }^2 \,p_{\parallel} }{4\,\pi }+3\,{b_d }^2 \,p_{\parallel} \,p_{\perp} \,{\left({b_d }^2 -2\right)}}{2\,\rho }.
\end{align*}

Note that the CGL wave speeds do not fully reduce to the MHD wave speeds when the parallel and perpendicular pressures are equal because of different adiabatic indices for the plasma in those two directions. The above set of eigenvalues ($\Lambda_d$) is valid within a specific admissible domain. To understand this physical domain, we first need to introduce two limits, which we call the lower limit (represented as \rev{$p_\mathrm{lo}$}) and the upper limit (represented as \rev{$p_\mathrm{hi}$}). These limits are defined as follows:
\begin{equation}
    p_\mathrm{lo} = \dfrac{p_\perp^2}{6p_\perp + \frac{3 B^2}{4 \pi}} \ \ ; \ \
    p_\mathrm{hi} = \dfrac{B^2}{4 \pi} + p_\perp.
    \label{eq:hyper_limits}
\end{equation}
The admissible domain, $\Omega$, to maintain hyperbolicity in the CGL system is given by
\begin{equation}
    \Omega=\{\mathbf{W} \in \mathbb{R}^9 : ~\rho>0, ~p_\parallel,p_\perp > 0,~p_\mathrm{lo} \leq p_{\parallel} \leq p_\mathrm{hi}  \},
    \label{eq:hyper_domain}
\end{equation}
where $\mathbf{W}=(\rho, \mathbf{v}, p_\parallel, p_\perp, \mathbf{B})^\top$ is the vector of primitive variables.
Unlike ordinary MHD, in the CGL system the slow wave is not necessarily slower than the Alfv\'en wave. Therefore, the admissible region for $p_\parallel$ can be divided into three parts \citep[see][]{kato1966propagation} based on the values of Alfv\'en wave ($c_A$), slow wave ($c_s$) and fast wave ($c_f$).
\rev{
\begin{enumerate}
    \item {$p_\mathrm{lo} \le p_\parallel \le \dfrac{p_\mathrm{hi}}{4}$},
          \qquad \qquad \qquad
          $c_s\le c_A \le c_f$
    \item {$\dfrac{p_\mathrm{hi}}{4} \le p_\parallel \le \dfrac{p_\mathrm{hi}}{4} + \dfrac{3p_\mathrm{lo}}{4}$},
          \qquad  \ 
          $c_s\le c_A < c_f$
    \item {$\dfrac{p_\mathrm{hi}}{4} + \dfrac{3p_\mathrm{lo}}{4} \le p_\parallel \le p_\mathrm{hi}$},
          \qquad  \  \
          $c_A\le c_s < c_f$.
\end{enumerate}
}

The complexity of the wave speeds and the realization that violating the bounds in the previous three equations could cause a code crash, leads us to an important realization: we will have to use physics to always keep guiding the numerics in such a way that the numerics can stay within the physically realizable domain. Eqn. ~\eqref{eq:press_jump} gives us an important clue that the relaxation time always drives the system towards isotropic pressure distributions. But for a given constant physical relaxation time, the system may not always stay out of the danger zone. In this paper, the above-mentioned goal is achieved by allowing the relaxation time to be modified when the pressure anisotropy becomes so large as to bring the system dangerously close to the limits of physical realizability. This is what we mean by an "elastic fence". The next section provides more detail.


\section{The physics of pressure relaxation} \label{sec:physics_of_relax}

This section is split into three subsections. The first subsection makes a physics-based case for having an ``elastic fence''. The second subsection provides the numerical details for implementing these ideas in code. The third subsection shows that this concept of an ``elastic fence'' also makes it possible for the CGL equations to attain an isotropic pressure in the non-magnetized limit.

\subsection{The physics-based case for an ``elastic fence''}

It was already recognized by \citet{1958PhRv..109.1874P} that a plasma with an anisotropic pressure distribution could become unstable and generate waves that would lead to a scattering of the particles and consequently drive the plasma toward isotropy. The CGL-MHD system predicts two kinds of instability; one operating in the regime $A<1$ (firehose) and the other requiring $A>1$ (mirror). The firehose instability appears on the Alfven wave branch. The region of stability is given by \citep{kato1966propagation}
\begin{equation}
\label{eq:firehose}
\beta_\perp>\beta_\parallel-2.
\end{equation}
The mirror instability, on the other hand, appears on the slow magnetosonic branch, for which the CGL stability criterion is \citep{kato1966propagation}
\begin{equation}
\label{eq:mirrorcgl}
\beta_\parallel>\frac{\beta_\perp^2}{6(1+\beta_\perp)}.
\end{equation}

Linear collisionless kinetic theory identifies three instabilities in a plasma with a bi-Maxwellian proton distribution. The firehose instability affects the Alfven-ion cyclotron (AIC) branch of the linear dispersion relation. It is strongest for parallel propagating low-frequency ($\omega\ll\Omega$) modes \citep{1990PhFlB...2..842Y, 2012JGRA..117.8101S}. The ion cyclotron instability is also associated with the AIC mode, but has higher frequencies ($\omega\sim\Omega$); its growth rate is maximized for parallel propagation. Finally, the mirror instability \citep{2012JGRA..117.8102Y} is a non-propagating compressive mode with maximum growth at oblique propagation. The mirror and AIC modes operate in a similar range of the $A$--$\beta$ space, for which reason they are sometimes grouped together in the literature despite being different in origin.

The kinetic instability thresholds are usually simplified to the limit of cold plasma, in which case the firehose criterion is identical to (\ref{eq:firehose}). The kinetic mirror instability has no factor of 6 in the denominator. If we delete that factor from (\ref{eq:mirrorcgl}) and solve for $\beta_\perp$, the expression becomes
\begin{equation}
\beta_\perp<\frac{\beta_\parallel}{2}+\sqrt{\beta_\parallel\left(1+\frac{\beta_\parallel}{4}\right)}
\label{eq:mirrorkin}
\end{equation}
Pressure anisotropies measured in the solar wind typically lie well inside the stability range and far from the marginal stability limits \citep{2006GeoRL..33.9101H}.

Expressions (\ref{eq:firehose}) and (\ref{eq:mirrorcgl}) correspond to the hyperbolicity conditions for the CGL system presented in Section \ref{sec:systemcgl}. As such, they cannot be violated, and a numerical model must perform an immediate correction to keep the system in the state of marginal stability. Note that there is no point in applying a relaxation term if the physical realizability criteria for the CGL system are not met because the model breaks down when the eigenvalues of the governing system of equations become complex. Physically, the imaginary part of the complex frequency (the growth rate) is comparable to the cyclotron frequency $\Omega$. The quasi-linear scattering rates are typically smaller, but still well outside of the time scales accessible to MHD models, so from that perspective the relaxation should be effectively instantaneous.

If the micro-instabilities were the only physical processes leading to a reduction in the anisotropy, then one would expect the plasma to often be in the state of marginal stability. However, the measured anisotropies are typically smaller than suggested by the stability criteria \citep{2004JGRA..109.4102M, 2006GeoRL..33.9101H}. In a large astrophysical system, such as the solar wind, the instabilities contribute to the turbulent spectrum of magnetic fluctuations at frequencies resonant with the particle's gyromotion. However, wavepower at relevant frequencies is also provided by the turbulent cascade from larger spatial scales. If the fluctuating power is not too large, the scattering frequencies could be much smaller than the cyclotron frequency and, therefore, comparable to the MHD timescales. The scattering rate can be estimated from the quasi-linear theory of interactions between particles and waves \citep{1966ApJ...146..480J, 1967PhFl...10.2620H, 1991A&A...250..266A}. The pitch-angle scattering coefficient is a complicated expression that depends on the pitch angle cosine $\mu$ and includes contributions from waves with multiple polarizations and directions of propagation. Very crudely, the expression for the turbulent scattering rate $\nu_\mathrm{scat}$ can be written as
\begin{equation}
\nu_\mathrm{scat}\simeq\frac{\Omega^2P(k_\mathrm{res})}{v^* B^2},
\label{eq:scatrate}
\end{equation}
where $v^*$ is the characteristic speed (the larger of the wave phase velocity, typically the Alfv\'en speed, and the particle thermal velocity), and the $P$ is the turbulent power spectral density at the resonant wavenumber $k_\mathrm{res}=\Omega/v^*$.

A typical core plasma particle is resonant in the dissipation turbulent range, where the spectral power is very small. Assuming a power law spectrum $P(k)\sim\langle\delta B^2\rangle l_c(k l_c)^{-\gamma}$, where $\langle\delta B^2\rangle$ is the mean square turbulent magnetic field magnitude, $l_c$ is the turbulent outer scale, and $v^*\sim v_t$, the thermal speed, one obtains
\begin{equation}
\nu_\mathrm{scat}\simeq\Omega\frac{\langle\delta B^2\rangle}{B^2} \left(\frac{r_L}{l_c}\right)^{\gamma-1},
\label{eq:scrate2}
\end{equation}
where $\langle\delta B^2\rangle/B^2$ is typically of the order of 0.1--1 (depending on the environment), and $r_L$ is the thermal Larmor radius. The spectral power index is typically between $-3/2$ and $-2$ in the inertial range and $-3$ in the dissipation range \citep{2016LNP...928.....B}. It is expected that as the \textit{mean field} $B\to 0$, the factor $\langle\delta B^2\rangle/B^2$ in expression (\ref{eq:scrate2}) becomes very large leading to a very efficient isotropization.

\begin{figure}[h]
	\begin{center}
		\includegraphics[width=0.98\textwidth,clip=]{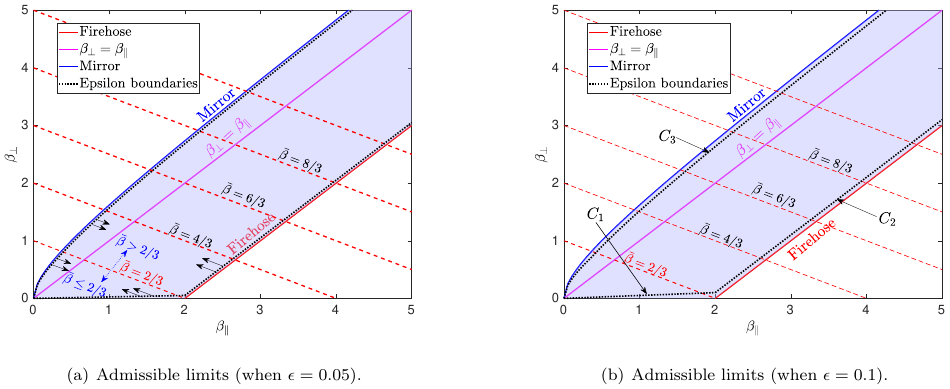}
        \caption{Domains of realizability of the CGL equations (shaded) for two values of $\epsilon$. The domain is most compactly expressed in terms of the plasma betas. The plasma betas for the parallel and perpendicular pressures form the horizontal and vertical axes of the plot. The magenta line shows the situation when the plasma pressures are isotropic. The mirror instability, shown by the thick blue line, forms the upper boundary of the domain of realizability for the CGL equations. The firehose instability, shown by the thick red line (as well as part of the x-axis), forms the lower boundary of the domain of realizability for the CGL equations. The dotted lines show slightly inner margins within which the plasma pressures should lie in a computation. The dashed red lines are lines of constant total pressure, and a small patch of plasma can only move along those lines if it is to conserve energy. (Of course, the energy of a patch of plasma is allowed to evolve with a simulation.) The arrows in Figure~\ref{fig:hyplim}a indicate directions along surfaces of constant total pressure.}
		\label{fig:hyplim}
	\end{center}
\end{figure}

\subsection{Numerical design of an ``elastic fence''}

Equation~\eqref{eq:scrate2} above gives us a physics-based measure of the relaxation time so that we can define a physical relaxation time as $\tau_{phys}  \equiv 1/\nu_{scat}$. Usually, for any physical system one has an estimate for the turbulent properties of the medium so that a physically-motivated relaxation time can be found. Even when the pressures are very close to isotropic, i.e. when \({p_\parallel } \simeq {p_ \bot }\), this relaxation time will be operational and will keep driving the pressures towards isotropization. However, in a computer code for modeling collisionless plasmas, it cannot be guaranteed that this rate of isotropization will always be sufficient to keep the system hyperbolic. Therefore, we develop a complimentary phenomenology based on the instability threshold expressions~\eqref{eq:firehose} and~\eqref{eq:mirrorkin} for the firehose and mirror instabilities, respectively. Figure~\ref{fig:hyplim}a illustrates the situation in the \({\beta _\parallel }\) versus \({\beta _ \bot }\) plane; and expressing the plasma pressures in terms of plasma beta is indeed the most compact way to show the results. The line of isotropy runs at a $45^\text{o}$ angle in Figure~\ref{fig:hyplim}a and is shown by the solid magenta line in Figure~\ref{fig:hyplim}a. The limit for the development of a firehose instability runs parallel to the magenta line and is shown by the red line in Figure~\ref{fig:hyplim}a. The development of the mirror instability is parallel to the $45^\text{o}$ line for large values of \({\beta _\parallel }\), with the result that asymptotically, i.e., for large values of \({\beta _\parallel }\), eqn.~\eqref{eq:mirrorkin} becomes \({\beta _ \bot } \le {\beta _\parallel } + 1\). Of course, for smaller values of \({\beta _\parallel }\), which can indeed occur when the magnetic field is strong, we will pay more careful attention to the permissible boundary for the mirror instability. The limit for the mirror instability is shown by the blue curve in Figure~\ref{fig:hyplim}a.

Now realize that the energy equation, i.e. eqns.~\eqref{eq:energy_conv} and~\eqref{eq:EOS}, will always permit us to use the code variables to extract the total pressure \(\bar p\) and the mean plasma beta \(\bar \beta \) which can be written as:- 
\begin{align}
\bar p \equiv {{\left( {{p_\parallel } + 2{p_ \bot }} \right)}/3} \Rightarrow \bar \beta  \equiv {{\bar p} /{\left( {{{\bf{B}}^2}/(8\pi )} \right)}}.
\label{eq:pbar}
\end{align}
To find independent values of \({p_\parallel }\) and \({p_ \bot }\) we have to resort to eqn.~\eqref{eq:press_jump} which gives us the evolution of \({p_\parallel } - {p_ \bot }\). This gives us two equations from which to disentangle \({p_\parallel }\) and \({p_ \bot }\). In practice, the two pressures should lie slightly inside the domain bounded by the mirror and firehose instabilities, as shown by the dotted curves in Figure~\ref{fig:hyplim}a. If they do not lie within the physically realizable domain at the end of a timestep in the code, energy should still be conserved with the result that we propose that the parallel and perpendicular pressures should move along the lines of constant total pressure till they do lie within the realizable domain. In other words the pressures move along the line \(\bar p \equiv {{\left( {{p_\parallel } + 2{p_ \bot }} \right)} / 3}\), which is equivalent to the dashed black lines \({\beta _\parallel } + 2{\beta _ \bot } = 3\bar \beta \) in Figure~\ref{fig:hyplim}a, until they lie within the realizable domain. Implicit in this realization is the fact that once the mirror or firehose instabilities are activated, they operate on plasma timescales that are shorter than the MHD timescales that govern the numerical timestep in the code. These fast-acting plasma instabilities re-establish a modicum of pressure isotropization, which will at least push the pressures within the physically realizable domain. Given that it is easier to work with plasma \(\beta 's\), for any given value of \(\bar \beta \) from the energy equation, we can find the limiting values of \(\beta _\parallel ^m\) and \(\beta _\parallel ^f\) for the mirror and firehose instabilities. They are given by 
\begin{align}
\beta _\parallel ^m = 2\left( {\bar \beta  + 1/3} \right) - \sqrt {{{(\bar{\beta}) }^2} + 8\bar \beta /3 + 4/9}
\label{eq:mirror_avg}
\end{align}
and
\begin{align}
\beta _\parallel ^f = \left\{ {\begin{array}{*{20}{c}}{\bar \beta  + 4/3{ \qquad \rm{for}} \qquad \bar \beta  \ge 2/3}\\{3\bar \beta {\qquad \qquad \rm{for    }} \qquad \bar \beta  < 2/3}\end{array}} \right.
\label{eq:firehose_avg}
\end{align}
The physically permissible values of \({\beta _\parallel }\) are given by \(\beta _\parallel ^m \le {\beta _\parallel } \le \beta _\parallel ^f\); please see Figure~\ref{fig:hyplim}a. This explains how the parallel and perpendicular pressures can always be brought within the physically realizable limits. The limits shown in Figure~\ref{fig:hyplim}a will also ensure that the system retains hyperbolicity as described in \S \ref{sec:eigenvalue}. 

\begin{figure}[h]
	\begin{center}
		\includegraphics[width=0.98\textwidth,clip=]{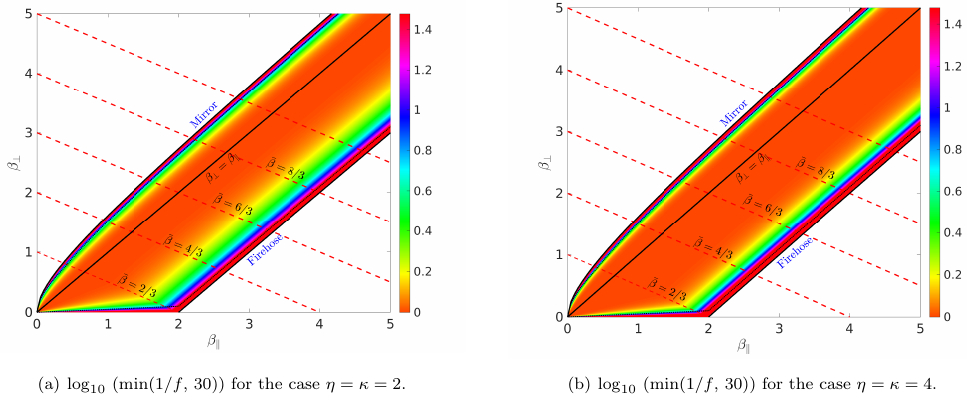}
		\caption{Panel shows the plot of $\log_{10}$ (min(1/$f$, 30)). We can see that the function is smooth even at the $45^\text{o}$ line of isotropy. It only assumes large values as we approach the boundaries of the mirror and firehose instabilities. This serves to keep the simulation within the realizable region. The choice of  the number “30” in the min function above is purely arbitrary and is intended to reveal the range of variation of “1/$f$”.}
		\label{fig:instab}
	\end{center}
\end{figure}

Since we understand how \({p_\parallel }\) and \({p_ \bot }\) can be kept within the bounds that ensure that the hyperbolic system is physically realizable, we now focus on obtaining a numerical relaxation time, \({\tau _{num}}\) which is suitable for numerical work. When it is numerically safe, \({\tau _{num}}\) should approach \({\tau _{phys}}\) quite closely. However, \({\tau _{num}}\) should be engineered so that it nudges the system away from the boundaries that are determined by the mirror and firehose instabilities in Figure~\ref{fig:hyplim}a. In other words, we seek a function \(f\left( {{\beta _\parallel },\bar \beta } \right)\) such that
\begin{align}
{\tau _{num}} = {\tau _{phys}}{\rm{ }}f\left( {{\beta _\parallel },\bar \beta } \right)
\label{eq:tau_num}
\end{align}
When \({\beta _\parallel } \to \beta _\parallel ^m\) or when \({\beta _\parallel } \to \beta _\parallel ^f\) we want the relaxation time to become very small; i.e., we want \(f\left( {{\beta _\parallel },\bar \beta } \right) \to 0\). In all other situations we want \(f\left( {{\beta _\parallel },\bar \beta } \right) \to 1\). Numerous choices for such a function are possible but the function should be such that when \({\beta _\parallel } < \bar \beta \), the plasma pressures may not hit the boundary formed by the mirror instability. Likewise, when \({\beta _\parallel } > \bar \beta \), the plasma pressures may not hit the boundary formed by the firehose instability. When \({\beta _\parallel } \to \bar \beta \) we want \(f\left( {{\beta _\parallel },\bar \beta } \right) \to 1\) so that the physics-based relaxation is achieved. To have maximal continuity, we want \(f\left( {{\beta _\parallel },\bar \beta } \right)\) to be a continuous function with zero slope at \({\beta _\parallel } = \bar \beta \). We see that it is acceptable to define \(f\left( {{\beta _\parallel },\bar \beta } \right)\) in a piecewise fashion as:-
\begin{align}
f\left( {{\beta _\parallel },\bar \beta } \right) 
= 
\left\{ 
{\begin{array}{*{20}{c}}{{{\cos }^\kappa }\left[ {\dfrac{\pi }{2}{{\left| {\dfrac{{\bar \beta  - {\beta _\parallel }}}{{\bar \beta  - \beta _\parallel ^m}}} \right|}^\eta }} \right]{ \qquad \rm{       when } \quad}{\beta _\parallel } < \bar \beta }\\{{{\cos }^\kappa }\left[ {\dfrac{\pi }{2}{{\left| {\dfrac{{{\beta _\parallel } - \bar \beta }}{{\beta _\parallel ^f - \bar \beta }}} \right|}^\eta }} \right]{\qquad \rm{when } \quad}{\beta _\parallel } > \bar \beta }\\{1 \qquad \qquad \qquad \qquad {\rm{otherwise}}}\end{array}} \right.
\label{eq:fbeta}
\end{align}
We can take \(\kappa  = 2\) or \(\kappa = 4\), and also \(\eta  = 2\) or \(\eta = 4\), in the above formula to make it increasingly flat around \({\beta _\parallel } = \bar \beta \). (Even and positive values of “\(\kappa \)” enable us to ensure that the derivatives of the above functions are zero at the limits over which they operate. Increasing “\(\kappa \)” increases the number of derivatives that are zero around \({\beta _\parallel } = \bar \beta \). Increasing “\(\eta \)”, while keeping it an even integer helps us to increase the range of values around \(\bar \beta \) over which the function \(f\left( {{\beta _\parallel },\bar \beta } \right)\) remains flat.) Figure~\ref{fig:instab} shows a colorized plot of the reciprocal of the above function in the \(\left( {{\beta _\parallel },{\beta _ \bot }} \right)\) plane. We see that as the plasma approaches the mirror or firehose instabilities, the reciprocal of the relaxation time becomes very large, i.e. the traversal away from the unrealizable part of the domain becomes very rapid. Otherwise, the numerical relaxation time should track the physical relaxation time, as evinced by unit values around the line of pressure isotropization in Figure~\ref{fig:instab}.

\subsection{The ``elastic fence'' concept allows the CGL Equations to reach the non-magnetic limit} \label{subsec:euler_lim}

One of the persistent concerns in using the CGL equations for numerical work is that the equations seem to become badly-formed in the limit where the magnetic field goes to zero. This was, indeed, one of the the motivating reasons behind the paper by~\cite{hirabayashi2016new}. We now show that the CGL equations, coupled with our new concept of an ``elastic fence" can smoothly produce an isotropic pressure in the limit where the magnetic field tends to (and reaches) zero in a numerical code.

In the limit when the magnetic field becomes very small, the plasma-\(\beta \) becomes very large. Therefore, we can obtain the asymptotic limits from eqns.~\eqref{eq:mirror_avg} and~\eqref{eq:firehose_avg}:
\begin{align}
    p_\parallel ^m = \bar p - \frac{2}{3}\left( {\frac{{{{\bf{B}}^2}}}{{8\pi }}} \right)
    \qquad
    ;
    \qquad
    p_\parallel ^f = \bar p + \frac{4}{3}\left( {\frac{{{{\bf{B}}^2}}}{{8\pi }}} \right)
\end{align}
We see, therefore, that \(p_\parallel ^m \le {p_\parallel } \le p_\parallel ^f\) . Furthermore, we see that the interval \(p_\parallel ^f - p_\parallel ^m = 2\left( {\dfrac{{{{\bf{B}}^2}}}{{8\pi }}} \right)\) shrinks with decreasing magnetic field. Therefore, as the magnetic field goes to zero, we are guaranteed to reach the limit \({p_\parallel } = {p_ \bot } = \bar p\). This ensures that we will asymptotically reach the limit of isotropic pressure as the magnetic field tends to zero. Notice that in that limit \({\bf{b}} = {{\bf{B}} \mathord{\left/
 {\vphantom {{\bf{B}} {\left| {\bf{B}} \right|}}} \right.
 \kern-\nulldelimiterspace} {\left| {\bf{B}} \right|}}\) can be any bounded unit vector that one wishes. The structure of the left-hand side of eqn.~\eqref{eq:press_jump} is such that any anisotropy that it induces during any fractional timestep is washed out at the end of that fractional timestep by our “soft fence” formulation. Furthermore, in the limit of zero magnetic field, eqn.~\eqref{eq:momentum_conv} will not depend on pressure anisotropy. Therefore, the hydrodynamical limit will always be reached. In other words, with all the advancements described here, the CGL equations will operate stably in the hydrodynamic limit because they reduce exactly to the MHD equations.

\section{Step by Step Implementation Plan}\label{sec:stepbystep}

\subsection{Nudging $\beta_\parallel$ and $\beta_\perp$ into the physical domain} \label{sec:nudging_to_physical}

The structure of the left-hand side of eqn.~\eqref{eq:press_jump} is such that it can always nudge the parallel and perpendicular pressures away from one another. Sometimes, if the terms on the left-hand side of eqn.~\eqref{eq:press_jump} are too large, it is always possible that the pressure anisotropy in a numerical code can become so large as to take the system past the bounds of physical realizability. In this section, we provide a method to nudge $\beta_\parallel$ and $\beta_\perp$ into the physical domain. We explicitly define the boundaries of the physical domain by introducing the curves $C_1$, $C_2$ and $C_3$ (which are pointed to by the black arrows in Figure~\ref{fig:hyplim}b). To ensure a computationally stable implementation, we maintain a positive distance ($\epsilon>0$) from the boundaries of the actual physical domain. This practice helps to prevent numerical errors that would arise in a numerical code if the solution were to approach the boundaries of Figure~\ref{fig:hyplim}b too closely. These 
 slightly shifted boundaries are shown with the dotted black lines in Figure~\ref{fig:hyplim}b. To make the dotted lines more conspicuous in Figure~\ref{fig:hyplim}b we have used $\epsilon=0.1$. In practice, we take $\epsilon=10^{-2}$ in all the numerical test problems. The curve $C_1$ denotes the lower boundary provided by the firehose instability with an added $\epsilon$-dependent buffer. The curve $C_2$ denotes the right boundary provided by the firehose instability with an added $\epsilon$-dependent buffer. The curve $C_3$ denotes the left boundary provided by the mirror instability with an added $\epsilon$-dependent buffer. With these notations, the required curves can be described as follows:
\begin{align*}
    C_1 \qquad &: \qquad \epsilon \beta_\parallel = 2 \beta_\perp,
    \\
    C_2 \qquad &: \qquad \beta_\parallel = (2+\beta_\perp) - \epsilon,
    \\
    C_3 \qquad &: \qquad \beta_\parallel = 
    \begin{cases}
    \dfrac{\beta_\perp^2}{(1+\beta_\perp)(1-\xi)}
    & 
    \text{if }
    \beta_\parallel  \le 2 - 2\sqrt{\frac{2}{3}}
    \\
    3\bar\beta - 2 \left(\beta_\perp^m - \epsilon \right)
    & 
    \text{if }
    \beta_\parallel  > 2 - 2\sqrt{\frac{2}{3}}
    \end{cases},
\end{align*}
where $\xi = \dfrac{((-6 + 4\sqrt{6} - 3\epsilon)\epsilon)}{(2 + 2(-3 + \sqrt{6})\epsilon)}$, $\beta_\perp^m = \dfrac{3\bar \beta - \beta _\parallel ^m}{2}$ and $\beta _\parallel ^m$ is given by eqn.~\eqref{eq:mirror_avg}.

%
We call the region enclosed within the dotted curves in Figure~\ref{fig:hyplim}b (i.e. the region surrounded by the curves $C_1$, $C_2$ and $C_3$) as the physically admissible domain, and we denote this domain by $\Omega_{\beta}$. We desire plasma beta's (i.e. $\beta_\parallel$ and $\beta_\perp$) to remain within the physical region at each grid point. To achieve this objective, we provide a step-by-step method to nudge $\beta_\parallel$ and $\beta_\perp$ into the physical domain.
\begin{enumerate}
    \item Realize that the energy equation, i.e. eqns.~\eqref{eq:energy_conv} and~\eqref{eq:EOS}, will always permit us to extract the total pressure \(\bar p\) and the mean plasma beta \(\bar \beta \) which can be written as:- 
    \begin{align}
    \bar p \equiv {{\left( {{p_\parallel } + 2{p_ \bot }} \right)}/3} \Rightarrow \bar \beta  \equiv {{\bar p} /{\left( {{{\bf{B}}^2}/(8\pi )} \right)}}.
    \label{eq:pbar_to_beta_bar}
    \end{align} 
    \item Eqn.~\eqref{eq:press_jump} gives us the evolution of \( \Delta p \equiv {p_\parallel } - {p_ \bot }\). This, along with eqn.~\eqref{eq:pbar_to_beta_bar}, gives us two equations which are solved to obtain the pressure variables \({p_\parallel }\) and \({p_\bot }\). 
    \item Utilizing \({p_\parallel }\) and \({p_\bot }\), we extract corresponding beta's (i.e. \({\beta_\parallel }\) and \({\beta_\bot }\)) using the following expressions:-
    \begin{align}
    \beta_\parallel  \equiv {{ p_\parallel} /{\left( {{{\bf{B}}^2}/(8\pi )} \right)}}
    \qquad ; \qquad
    \beta_\perp  \equiv {{ p_\perp} /{\left( {{{\bf{B}}^2}/(8\pi )} \right)}}
    \label{eq:beta_s}
    \end{align} 
    \item Realize that the obtained plasma beta's, i.e. $\beta_\parallel$ and $\beta_\perp$  in eqns.~\eqref{eq:beta_s}, may have been nudged out of the physical domain $\Omega_\beta$. We need to nudge $\beta_\parallel$ and $\beta_\perp$ back into the physical domain in a thermal energy-conserving fashion to retrieve the physically admissible plasma beta's $\hat \beta_\parallel$ and $\hat \beta_\perp$.
    \item If $\beta_\parallel$ and $\beta_\perp$ have stepped out of the physical domain, the following expressions for $\hat \beta_\parallel$ and $\hat \beta_\perp$ will nudge them into the admissible domain $\Omega_\beta$.
    \begin{align}
        \hat \beta_\parallel =&
        \max \left\{ 
        \min \left\{
        \alpha_1,
        \beta_\parallel
        \right\},
        3 \bar\beta - 2 \alpha_2
        \right\}, \qquad ; \qquad \hat \beta_\perp = \dfrac{3 \bar\beta - \hat \beta_\parallel}{2}
        \label{eq:phys_beta_par}
        \intertext{where}
        \alpha_1&=\min \left\{
        \dfrac{3 \bar\beta}{\epsilon+1}, \bar\beta + \dfrac{2}{3}(2-\epsilon)
        \right\} \quad;\quad \alpha_2 = \max \left\{
        \dfrac{\left(-\text{b}+\sqrt{\text{b}^2-\text{4ac}}\right)}{\text{2 a}}
        ,
        \beta_\perp^m - \epsilon 
        \right\} \ ;
        \nonumber
        \\
        \text{a} &=(3-2\xi)\ ; \ \text{b} =(1-\xi)(2-3\bar \beta)\ ; \  \text{c} =-3 (1-\xi) \bar \beta
        \nonumber \quad ;
        \\
        \xi &= \dfrac{(-6 + 4\sqrt{6} - 3\epsilon)\epsilon}{2 + 2(-3 + \sqrt{6})\epsilon}\ ; \ \beta_\perp^m = \dfrac{3\bar \beta - \beta _\parallel ^m}{2} \ ; \ \beta _\parallel ^m = 2\left( {\bar \beta  + 1/3} \right) - \sqrt {{{(\bar{\beta}) }^2} + 8\bar \beta /3 + 4/9}.
        \nonumber
    \end{align}
    Realize that if $\beta_\parallel$ and $\beta_\perp$ are already inside the physical domain $\Omega_\beta$ then the expressions in eqn.~\eqref{eq:phys_beta_par} are designed to identically retrieve the beta's given in~\eqref{eq:beta_s}; this ensures that the expressions in eqn.~\eqref{eq:phys_beta_par} do not change the beta's in eqn.~\eqref{eq:beta_s} if the latter are already within the physical domain.
    \item We replace the expressions $\beta_\parallel$ and $\beta_\perp$ given in~\eqref{eq:beta_s} with the physically admissible beta's $\beta_\parallel$ and $\beta_\perp$ given in eqn.~\eqref{eq:phys_beta_par} (i.e., we reset $\beta_\parallel = \hat \beta_\parallel$ and $\beta_\perp = \hat \beta_\perp$). This in turn yields physically realizable parallel and perpendicular pressures that can be reset in the code. Notice that our formulation preserves the mean thermal energy in the plasma. In other words, the $\hat \beta_\parallel$ and $\hat \beta_\perp$ lie on the same red dashed curves in Figure~\ref{fig:hyplim}b as the original $\beta_\parallel$ and $\beta_\perp$.
    \item When the magnetic field is below some threshold that the applications scientist could view as zero, then we can use the logic from Sub-section~\ref{subsec:euler_lim} to justify the setting. The same consideration applies to our evaluation of eqn.~\eqref{eq:fbeta}.
\end{enumerate}
    The above seven-step procedure can always be employed whenever one wants to go from conserved to primitive variables in a numerical code. It ensures that the numerical solution always lies within the domain of physical realizability.
    
\subsection{Relaxation times: Physical and Numerical consideration} \label{sec:relaxation}

While the previous Sub-section shows us how to always keep the solution physically realizable, we need to describe the ``elastic fence'' and explain why it works. We will always have some observationally motivated, or numerically generated value for the turbulence in the system. Call the timescale for those turbulent eddies $\tau _{phys}$. In principle, we would like to insert this value of $\tau _{phys}$ in eqn.~\eqref{eq:press_jump} and hope that the difference in pressures always stays within the domain of physical realizability. But that hope is almost never respected by a numerical code. Having a bounds-preserving timestepping strategy will help. However, when the pressure anisotropy gets too close to the ``hard fence'' formed by physical realizability, it is valuable to provide instead an ``elastic fence'' which keeps the pressure anisotropy from touching, or going beyond, the ``hard fence''. This is provided by eqn.~\eqref{eq:tau_num}, which in turn relies on eqns.~\eqref{eq:mirror_avg}, ~\eqref{eq:firehose_avg} and ~\eqref{eq:fbeta}. Collectively, these equations ensure that the relaxation time is made much shorter than $\tau _{phys}$ when the system gets too close to the boundaries of the domain of realizability.

\begin{figure}[h]
\begin{center}
\includegraphics[width=0.5\textwidth,clip=]{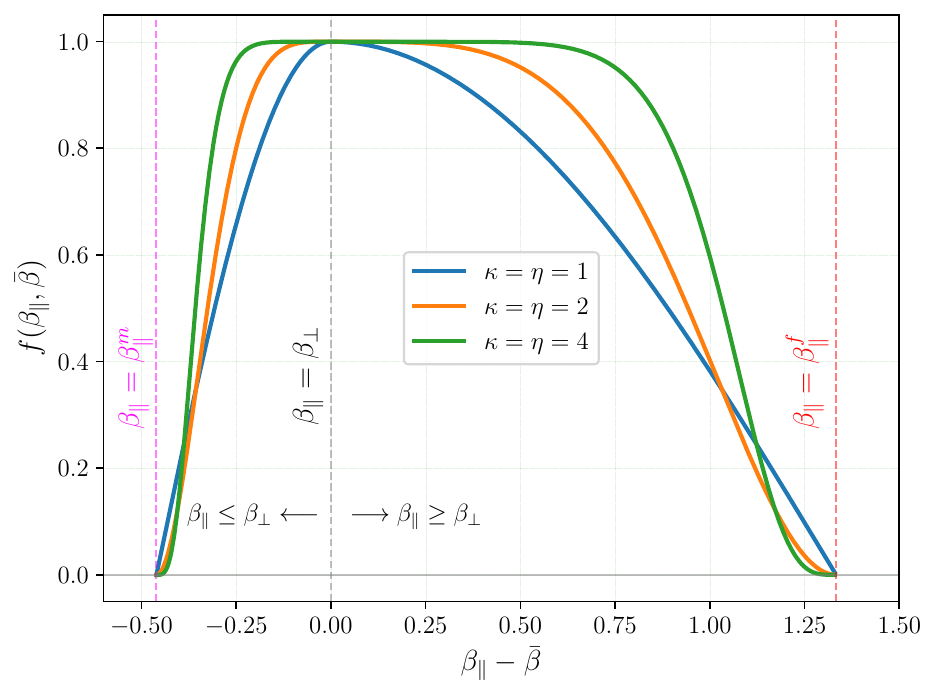}
\caption{Graph of $f(\beta_\parallel,\bar\beta)$ for $\bar\beta=6.0/3.0$.}
\label{fig:f_beta_1d}
\end{center}
\end{figure}

Fig.~\ref{fig:f_beta_1d} shows how this works. In this figure, we have set $\bar\beta = 6/3$. To conserve total thermal energy, the values of $\beta_\parallel$ and $\beta_\perp$ must move along the red dashed line given by $\bar\beta = 6/3$ in Fig.~\ref{fig:hyplim}b. We now plot out $f\left( {{\beta _\parallel },\bar \beta } \right)$ for that situation in Fig.~\ref{fig:f_beta_1d}. We see that for a large range of values, $f\left( {{\beta _\parallel },\bar \beta } \right)$ is close to unity, with the result that the relaxation time is indeed given by $\tau _{phys}$. When $\beta _\parallel$ is much greater than $\bar \beta$, or when it is much smaller than $\bar \beta$ , i.e., when it approaches either asymptote in Fig.~\ref{fig:f_beta_1d}, we see that $f\left( {{\beta _\parallel },\bar \beta } \right)$ approaches zero. This causes $\tau _{num}$ to become a very small value. In turn, this causes the numerical relaxation time in eqn.~\eqref{eq:press_jump} to become very small and the system is made to move away from the ``hard fence'' before the numerical code loses saliency. Please also note from Fig.~\ref{fig:f_beta_1d} that we want $f\left( {{\beta _\parallel },\bar \beta } \right)$ to remain close to unity for a large range of values. This is ensured if we take \(\kappa  = 2\) or \(\kappa = 4\), and also \(\eta  = 2\) or \(\eta = 4\) in eqn.~\eqref{eq:fbeta}. We also see that setting \(\eta  = 1\) and \(\kappa  = 1\) is a sub-optimal choice. As long as the prescription from the previous Sub-section is used, $\beta _\parallel$ will never be equal to the dashed asymptotic values shown in Fig.~\ref{fig:f_beta_1d}, so it will never become exactly zero. For all the examples shown in this paper, we have used \(\eta  = 4\) and \(\kappa  = 4\).

\subsection{Overall Implementation} \label{sec:implementation}
The previous two Sub-sections have shown us how to ensure two things. First, Sub-section~\ref{sec:nudging_to_physical} has shown us that ${{p}_{\parallel }}$ and ${{p}_{\bot }}$ can always be made to lie in the physical region, even if perchance the numerical errors may take them out of the physical region. This should occur very, very rarely in a code. Second, Sub-section~\ref{sec:relaxation} showed that we have a physics-based relaxation strategy that always ensures that under normal operation the numerics remain safely within the region of realizability. For the timestepping we will use one of the two Runge-Kutta schemes described in subsection~\ref{sec:PDE_time_stepping}. Here, we provide a step-by-step implementation plan for a single stage in the Runge-Kutta timestepping:-
\begin{enumerate}
    \item Start with the zone-centered conserved variables and face-centered magnetic field variables. The zone centered variables are reconstructed using finite volume WENO methods; see for example, the Supplement for \cite{balsara2023efficient}. The divergence-free magnetic field can be reconstructed using the methods described in \cite{balsara2018computational}. 
    \rev{(Any reasonable TVD or WENO scheme can be used for the reconstruction of zone-centered fluid variables at second order. At third order, we reconstructed the zone-centered fluid variables by using Sub-section A.1 from the Supplement in~\cite{balsara2023efficient}. For the divergence-free magnetic field, we used the second order reconstruction from Section III.2 of~\cite{balsara2001divergence}. At third order, we reconstructed the magnetic field by using Section 2 of~\cite{balsara2009divergence} or Section 3 of~\cite{balsara2018computational}.)}
    \item The reconstruction from the previous step gives us non-linearly hybridized polynomials for the variation of all the flow variables within each zone. Now focus on Fig.~\ref{fig:grid}. Using the reconstructed variables within each zone, we can use those polynomials to obtain the nodal values within each zone. We can also use those nodal values to evaluate the fluxes at each nodal point. This enables us to evaluate the flux terms, which form the third and fourth rows of eqn.~\eqref{eq:Lupdate}. This is done for each zone. We can also obtain the gradient of the solution at each nodal point. Along with the matrices for the non-conservative products, we can use the gradients to evaluate the fifth row of eqn.~\eqref{eq:Lupdate}. This is done for each zone.
    \item As a result of the nodal evaluation in the previous step, at each zone boundary we have two nodal values on either side of that zone boundary. These can be fed into the one-dimensional Riemann solver from~\cite{dumbser2016new} to get fluctuations that propagate to either side of each zone boundary. This allows us to evaluate the first two rows of eqn.~\eqref{eq:Lupdate}. We see, therefore, that the entire time rate of change for the zone-centered variables, as given by eqn.~\eqref{eq:Lupdate}, is now in hand. 
    \item Also as a result of the nodal evaluation in step 2, we have four states at each edge. These are fed into a two-dimensional Riemann solver to get electric fields at each edge. \cite{balsara2010multidimensional,balsara2012two,balsara2014multidimensional} or \cite{balsara2017multidimensional} describe such multidimensional Riemann solvers. The resulting electric fields at the edges of the mesh can then be used to obtain the time rate of change for the facial magnetic field components. 
    \item This completes our description of how $\mathcal{L}(\bf{U})$ is obtained for use in eqn.~\eqref{eq:timeo2} or eqn.~\eqref{eq:timeo3}. \rev{(Eqn.~\eqref{eq:timeo2} will provide a temporally second order accurate update; eqn.~\eqref{eq:timeo3} will provide a temporally third order accurate update.)} Note too that at the beginning of each complete timestep we evaluate $\tau_{num}$ within each zone using eqns.~\eqref{eq:tau_num} and~\eqref{eq:fbeta}. This $\tau_{num}$ can differ from zone to zone, and it can indeed change from one complete timestep to another, but it is kept frozen within each zone for the duration of an entire timestep. This is so that it can be used in eqn.~\eqref{eq:timeo2} or eqn.~\eqref{eq:timeo3}, which only remain salient if $\tau_{num}$ is kept frozen within each zone for the duration of that timestep. 
\end{enumerate}

\section{Accuracy tests} \label{sec:accuracytests}
In~\cite{balsara2004second}, a two-dimensional vortex problem was introduced within the framework of the MHD system. In this section, we design such a test problem for the CGL system. This problem is especially valuable for accuracy testing due to its unique characteristics, featuring a smoothly evolving and dynamically stable vortex that moves diagonally within the computational domain.

The problem is defined within a two-dimensional domain that spans the region $[-5,5]\times[-5,5]$. Periodic boundaries have been implemented at all the boundaries. The vortex is initiated at the center of the computational domain through fluctuations in the velocity and magnetic fields. These fluctuations are defined as follows:
\begin{align}
    (\delta v_x, \delta v_y)
    =
    \zeta \ e^{0.5(1-r^2)}(-y,x),  \label{eqn:vel}
    \\
    (\delta B_x, \delta B_y)
    =
    \mu \ e^{0.5(1-r^2)}(-y,x), \label{eqn:B}
\end{align}
where $\zeta$ and $\mu$ are positive parameters that will be defined later. The distance from the center of the domain is denoted by $``r"$, which is defined as the square root of the sum of squares of the coordinates, i.e., $r=\sqrt{x^2+y^2}$. It is important to emphasize that the magnetic vector potential plays a crucial role in ensuring that the magnetic field is initialized in a divergence-free manner within the computational domain. The expression for the magnetic vector potential, denoted as $A_z$, is given by
\begin{equation}
    A_z = \mu \ e^{0.5(1-r^2)}.
\end{equation}
As in the original MHD vortex from ~\cite{balsara2004second}, the unperturbed density, pressures, and velocities are taken to be constant. Moreover, for a given constant $\sigma$, we assume the following pressure relation for the perturbations $\delta p_\parallel$ and $\delta p_\perp$
\begin{equation}
    \delta p_\parallel - \delta p_\perp
    =
    \sigma^2 \ e^{(1-r^2)} r^2. \label{eqn:prelation} 
\end{equation}
For the CGL system, the equation for the dynamical balance is given by
\begin{align}
    \frac{\partial}{\partial r}
    \left(
        \delta p_\perp + \frac{{B}^2}{8 \pi}
    \right)
    =
    \frac{1}{r}
    \left(
         \rho {v}^2 + \delta p_\parallel - \delta p_\perp -\frac{{B}^2}{4 \pi}
    \right).
    \label{eqn:dynamical_balance}
\end{align}
Assuming that $\delta p_\perp$ is zero at the center of the domain, we solve equation~\eqref{eqn:dynamical_balance} to obtain
\begin{align}
    \delta p_\perp(r) 
    =
    \frac{\mu^2}{8\pi} \ (1-r^2)  \ e^{(1-r^2)} 
    - \frac{\mu^2}{8\pi} e
    -\frac{1}{2}
    \left(
         \rho \zeta^2 + \sigma^2
    \right)
    \left(
         e^{1-r^2} - e
    \right).
\end{align}
Consequently, using equation~\eqref{eqn:prelation}, we find
\begin{align}
    \delta p_\parallel = \delta p_\perp
    +
    \sigma^2 \ e^{(1-r^2)} r^2.
\end{align}
As in the MHD vortex, the unperturbed density and pressures ($p_\perp$ and $p_\parallel$) are set to unity. The unperturbed $x$ and $y$ velocities are set to unity, and the unperturbed magnetic field is set to zero. No perturbations have been applied to the density variable. Collectively, we have
\begin{equation}
    \left(\rho,v_x,v_y,v_z,p_\parallel,p_\perp,B_x,B_y,B_z\right)
    =
    \left(1,1+\delta v_x,1+\delta v_y,0,1+\delta p_\parallel,1+\delta p_\perp,\delta B_x,\delta B_y,0\right).
\end{equation}
The test problem was configured with the following parameters: $\mu = \sqrt{4\pi}/(2\pi)$, $\zeta = 1/(2\pi)$, and $\sigma = 1/(4\pi)$. We remark that setting $\sigma = 0$ retrieves analytic expressions of the MHD vortex problem that are available in~\cite{balsara2004second}. We simulate the vortex problem till time $t=10$ on different grid stacks to perform the accuracy study. Table~\eqref{table:accuracy} shows the obtained $L_1-$ and $L_\infty-$errors for the $B_y-$component of the magnetic field $\mathbf{B}$. The corresponding numerical orders are also given in the same table. For the fourth-order scheme, we double the computational domain and stopping time to minimize the effect of small jumps in the velocity and magnetic field at the periodic boundaries. We see that all the schemes (second, third, and fourth-order) reach their design accuracy.
\begin{table}[h]
	\centering
	\begin{tabular}{|c||c|c|c|c|c|}
		\hline
		Order of the scheme  & Grid size & $L_1-$errors & Order & $L_\infty-$errors & Order \\
           \hline
           \hline
		\multirow{4}{*}{Second order}
           & $32  \times 32  $ & 1.45602E-02 &  --    & 2.26862E-01 &   -- \\
           & $64  \times 64  $ & 4.04014E-03 &  1.85	& 5.94939E-02 &  1.93 \\
           & $128 \times 128 $ & 9.14568E-04 &  2.14	& 1.61236E-02 &  1.88 \\
           & $256 \times 256 $ & 2.25164E-04 &  2.02	& 4.54362E-03 &  1.83 \\
           \hline
		\multirow{4}{*}{Third order}
           & $32  \times 32  $ & 6.98221E-03	 &   --  & 1.15328E-01 &	--  \\
           & $64  \times 64  $ & 1.42498E-03	 & 2.29	& 2.28921E-02 &  2.33 \\
           & $128 \times 128 $ & 2.04612E-04	 & 2.80	& 4.73716E-03 &  2.27 \\
           & $256 \times 256 $ & 2.74924E-05	 & 2.90	& 7.50811E-04 &  2.66 \\
           \hline
		\multirow{4}{*}{Fourth order}
           & $32  \times 32  $ & 5.40291E-03 &  --    & 2.97769E-01 &    --  \\
           & $64  \times 64  $ & 6.34923E-04 &  3.09	& 4.87071E-02 &  2.61 \\
           & $128 \times 128 $ & 2.81249E-05 &  4.50	& 2.99674E-03 &  4.02 \\
           & $256 \times 256 $ & 1.70317E-06 &  4.05   & 1.95570E-04 &  3.94 \\
           \hline
	\end{tabular}
	\caption{\nameref{sec:accuracytests}: $L_1$ errors, $L_\infty$ errors, and the corresponding order of convergence for the second, third, and fourth-order schemes.}
	\label{table:accuracy}
\end{table}

\section{Some example problems} \label{sec:prob}

In this Section, we upgrade several standard two-dimensional MHD test problems to show their corresponding CGL variants. For schemes characterized by spatial second-order accuracy, temporal evolution is governed by a second-order time-stepping algorithm specified in eqn.~\eqref{eq:timeo2}. The schemes exhibiting spatial third and fourth-order accuracy employ a third-order time-stepping strategy detailed in eqn~\eqref{eq:timeo3}. For all the simulations presented in this section, we used a CFL number of 0.4. 

\subsection{Rotor Problem for the CGL Equations} \label{subsec:rotor}

\begin{figure}[!ht]
	\begin{center}
		\includegraphics[width=0.90\textwidth,clip=]{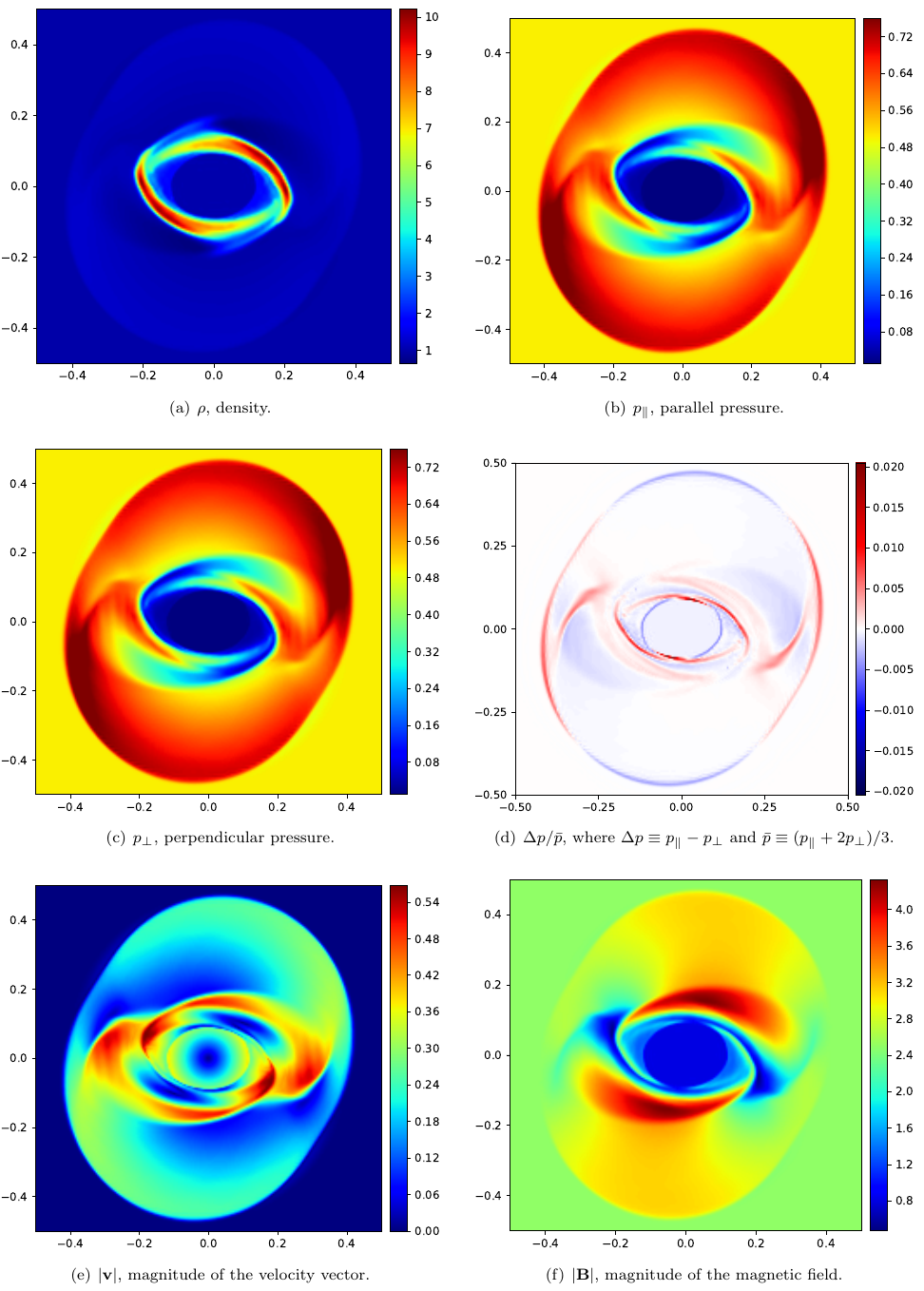}
		\caption{\nameref{subsec:rotor}: The panels show the plots of $\rho$, $p_\parallel$, $p_\perp$, $\Delta p / \bar p$, $|\bf{v}|$ and $|\bf{B}|$ that correspond to $\tau _{phys} = 10^{-5}$.}
		\label{fig:rotor1}
	\end{center}
\end{figure}

\begin{figure}[!ht]
	\begin{center}
        \includegraphics[width=0.90\textwidth,clip=]{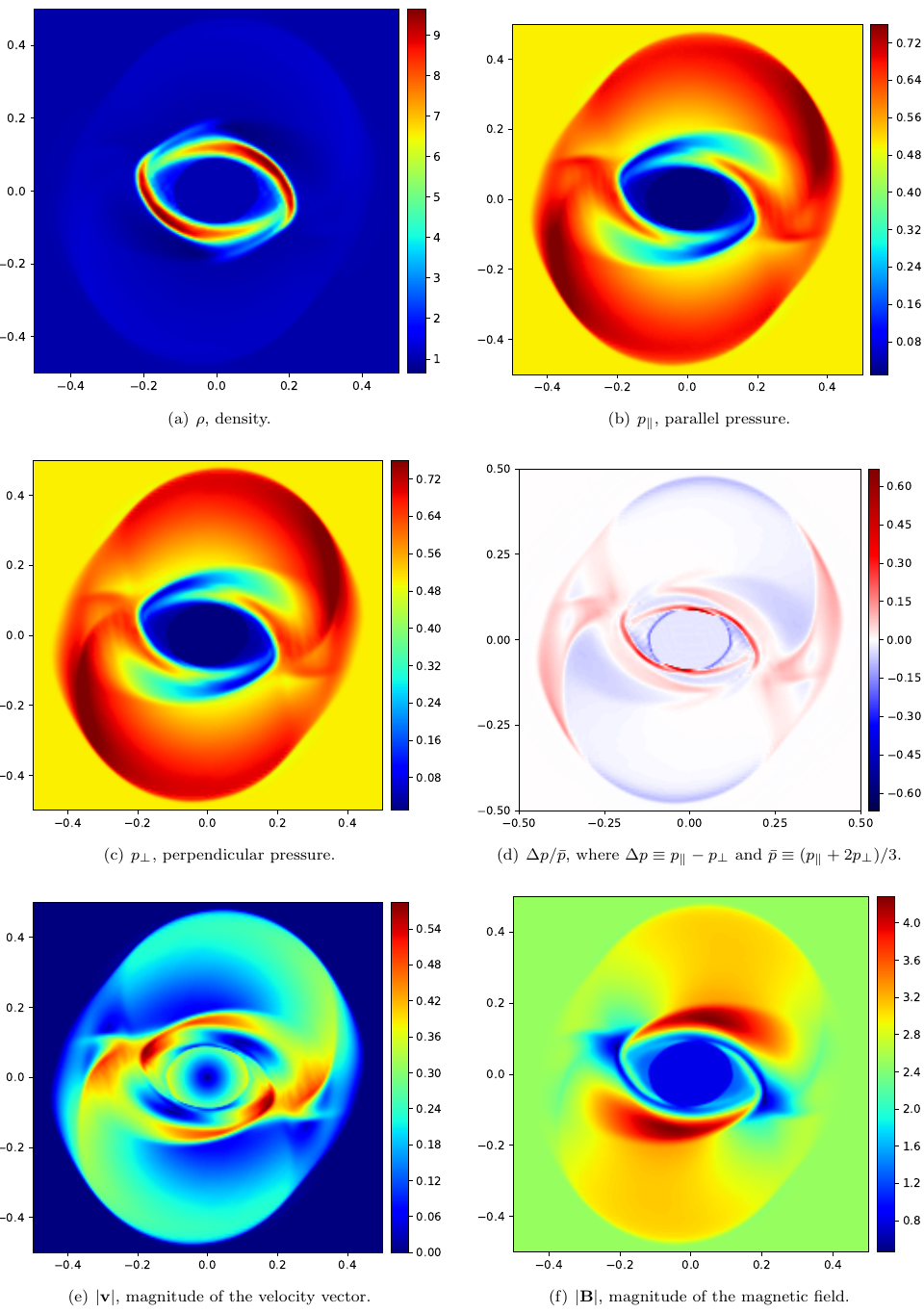}
		\caption{\nameref{subsec:rotor}: The panels show the plots of $\rho$, $p_\parallel$, $p_\perp$, $\Delta p / \bar p$, $|\bf{v}|$ and $|\bf{B}|$ that correspond to $\tau _{phys} = 10^{-2}$.}
		\label{fig:rotor2}
	\end{center}
\end{figure}

\begin{figure}[!ht]
	\begin{center}
		\includegraphics[width=0.90\textwidth,clip=]{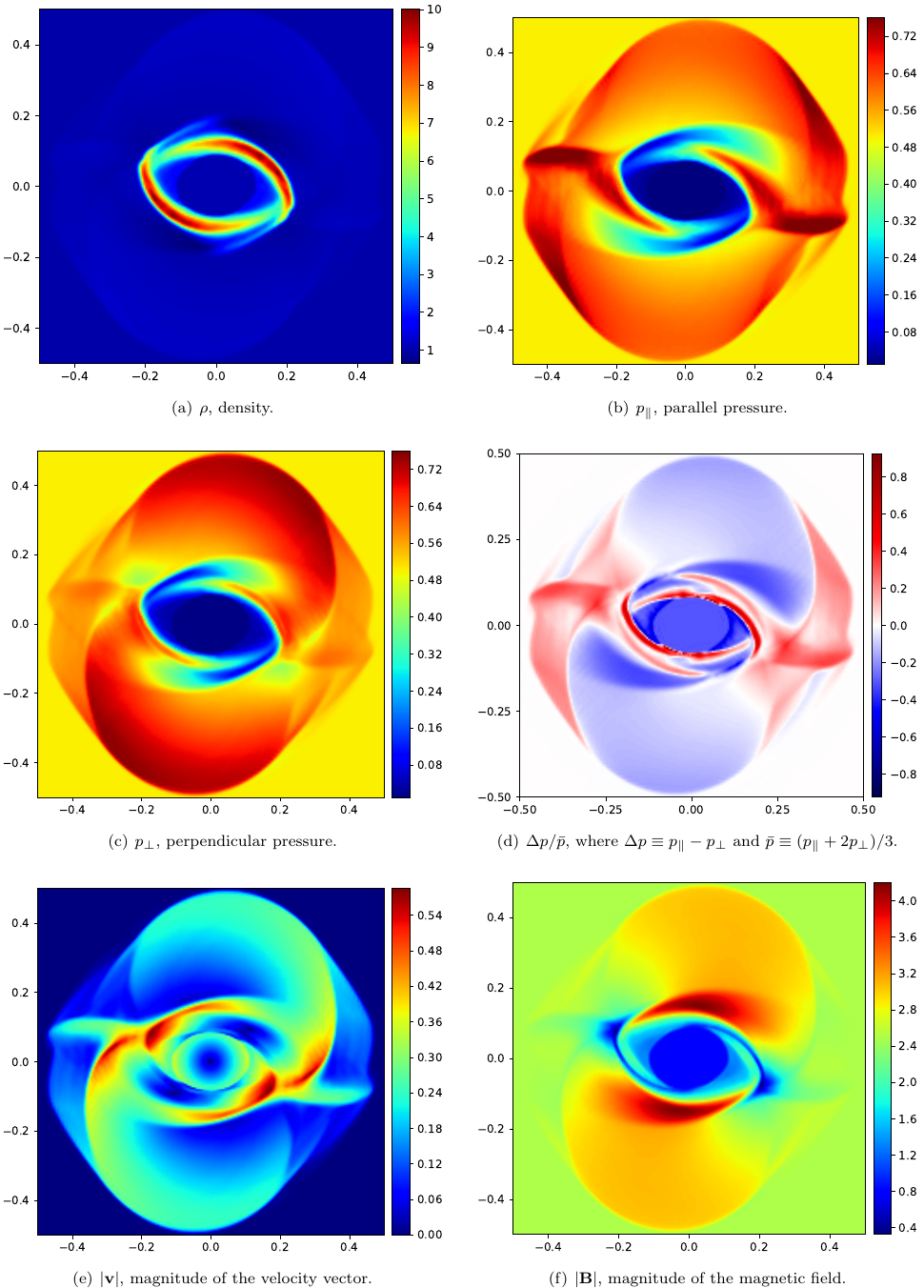}
		\caption{\nameref{subsec:rotor}: The panels show the plots of $\rho$, $p_\parallel$, $p_\perp$, $\Delta p / \bar p$, $|\bf{v}|$ and $|\bf{B}|$ that correspond to $\tau _{phys} = 0.1$.}
		\label{fig:rotor3}
	\end{center}
\end{figure}

In this sub-section, we consider a CGL version of the two-dimensional rotor problem that was first presented for MHD in~\cite{balsara1999staggered} and~\cite{balsara2004second}. The computational domain spans the domain $[-0.5,0.5]\times[-0.5,0.5]$. A dense and rapidly spinning cylinder is set up in the center of an initially stationary, light ambient fluid. The ambient fluid is initially at rest. A uniform magnetic field initially threads the two fluids. Its value is set to 2.5 units, and it initially points in the x-direction. The total pressure in both fluids is set to unity, i.e. $\bar p= 1.0$. The initial parallel and perpendicular pressures are assumed to be equal to the total pressure, i.e. $p_\parallel=p_\perp=\bar p$. The density in the ambient fluid is uniformly set to unity, while the constant density in the rotor is $10$ units out to a radius of $0.1$. A linear taper is applied to the density between a radius of $0.1$ and $0.13$ so that the density in the rotor decreases linearly to the value of the density in the ambient fluid. Six zones are used for the taper to join the density of the two fluids. That number should be kept fixed if the resolution is increased or decreased. The initial angular velocity of the rotor is uniform out to a radius of $0.1$. At this radius, the toroidal velocity has a value of one unit. The toroidal velocity decreases linearly from one unit to zero between a radius of 0.1 and 0.13 so that it joins the velocity of the ambient fluid at a radius of 0.13. The simulations were run until a final time of $t=0.29$ using a mesh consisting of $200\times200$ zones. We use the fourth-order accurate scheme for this test problem and show the results for various $\tau_{phys}$ in Figs.~\ref{fig:rotor1}-\ref{fig:rotor3}.

Fig.~\ref{fig:rotor1} corresponds to $\tau _{phys} = 10^{-5}$ with the result that our physical relaxation time is substantially smaller than the timestep in this problem. As a result, we see that the pressures relax to their isotropic values within one timestep; so the MHD limit is practically reached. Consequently, we see that this simulation looks very much like the classical MHD case with an isotropic pressure. Fig.~\ref{fig:rotor1}d also shows that the relative anisotropy in the pressures is minimal. Fig.~\ref{fig:rotor2} corresponds to $\tau _{phys} = 10^{-2}$ with the result that our physical relaxation time is much larger than the timestep in this problem. We expect to see substantial differences in $p_\parallel$ and $p_\perp$, and Fig.~\ref{fig:rotor2}d for the relative anisotropy in the pressures shows a much larger variation than Fig.~\ref{fig:rotor1}d. Fig.~\ref{fig:rotor3} corresponds to $\tau _{phys} = 0.1$ with the result that our physical relaxation time is comparable to the final simulation time in this problem. Now Fig.~\ref{fig:rotor3}d shows extreme anisotropies in the pressure; please compare it to Fig.~\ref{fig:rotor2}d and Fig.~\ref{fig:rotor1}d. 

In this problem, the parallel and perpendicular pressures are identical at the start of the simulation and the magnetic field is initially horizontal. We see that the pressure anisotropies have been generated spontaneously by the governing CGL equations. In Sub-section~\ref{sec:intro_to_plasma}, we established some scalings for the parallel and perpendicular temperatures, which can also be extended to refer to scalings for the parallel and perpendicular pressures. By looking at the top and bottom of the rotor in Figs.~\ref{fig:rotor2}c and~\ref{fig:rotor3}c, we see that the outward propagating wave is propagating into the horizontal field, causing a substantially greater increase in the perpendicular pressure compared to the parallel pressure. However, we can also focus on the horizontal ends of the rotor in Figs.~\ref{fig:rotor2}b and~\ref{fig:rotor3}b, which show the parallel pressure. There, we see that the rotation is stretching the field lines longitudinally with the result that the parallel pressure has increased much faster than the perpendicular pressure.

\subsection{Blast Problem for the CGL Equations}\label{subsec:blast}

\begin{figure}[!ht]
	\begin{center}
		\includegraphics[width=0.90\textwidth,clip=]{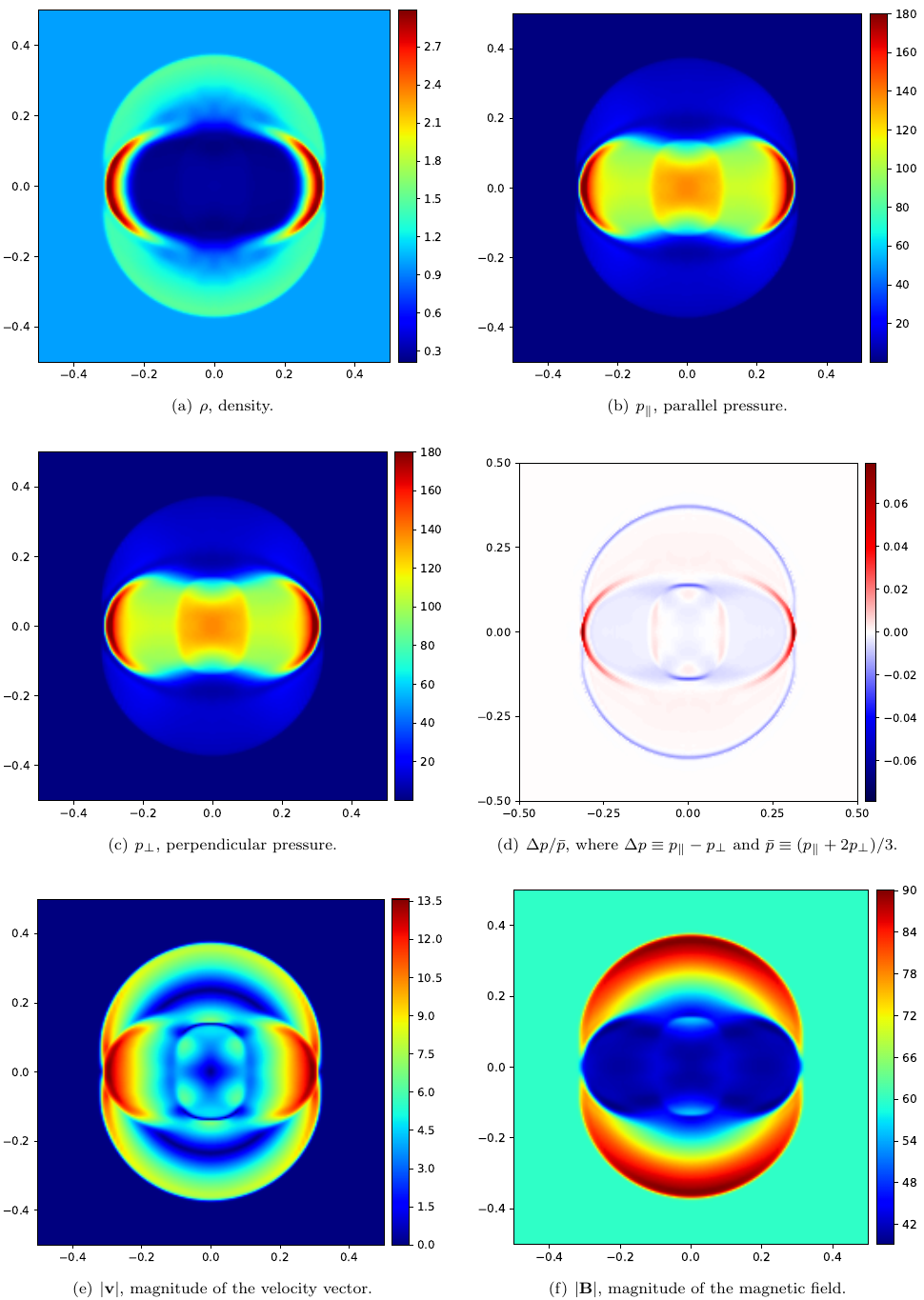}
		\caption{\nameref{subsec:blast}: The panels show the plots of $\rho$, $p_\parallel$, $p_\perp$, $\Delta p / \bar p$, $|\bf{v}|$ and $|\bf{B}|$ that correspond to $\tau _{phys} = 10^{-5}$.}
		\label{fig:blast1}
	\end{center}
\end{figure}

\begin{figure}[!ht]
	\begin{center}
		\includegraphics[width=0.90\textwidth,clip=]{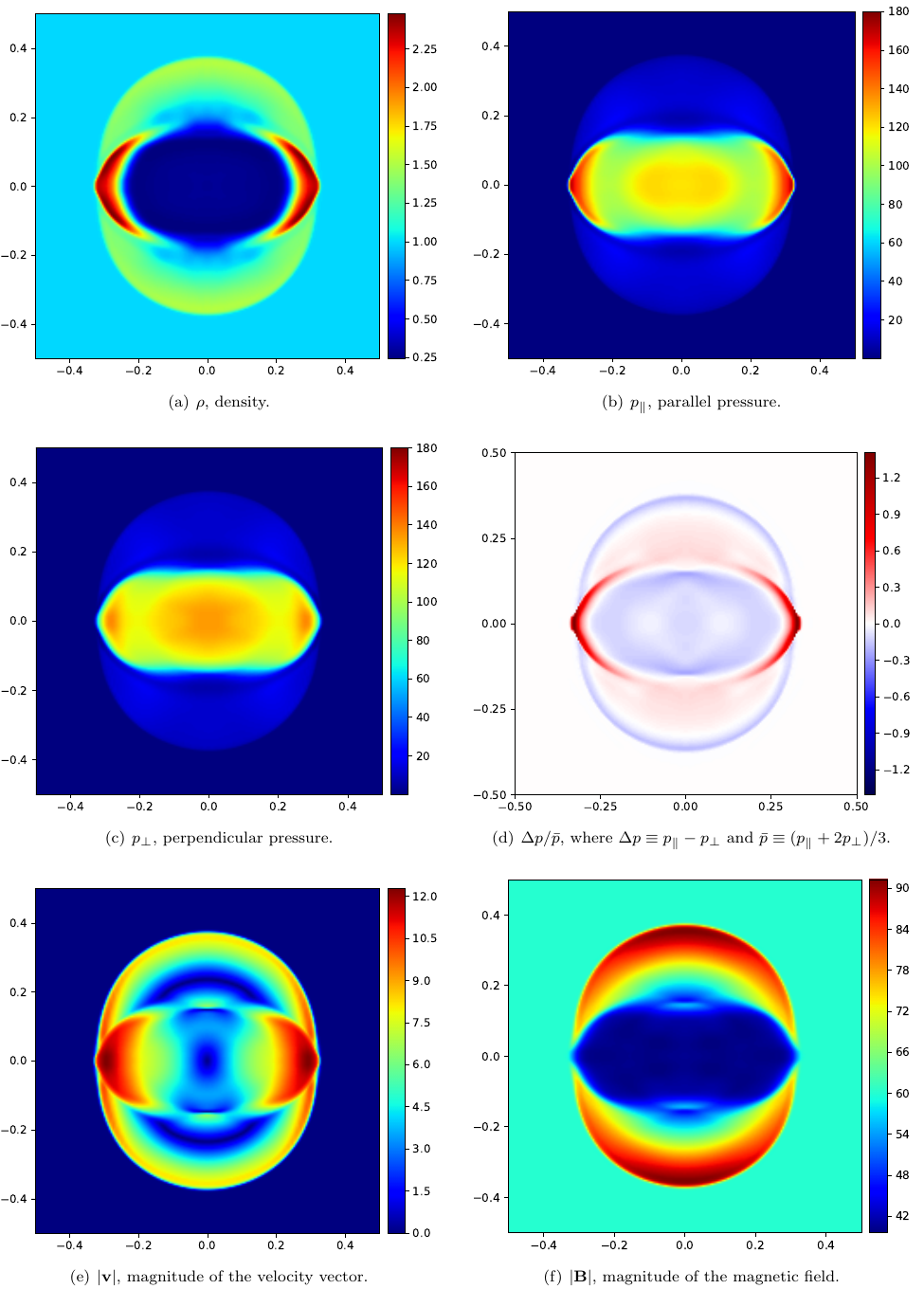}
		\caption{\nameref{subsec:blast}: The panels show the plots of $\rho$, $p_\parallel$, $p_\perp$, $\Delta p / \bar p$, $|\bf{v}|$ and $|\bf{B}|$ that correspond to $\tau _{phys} = 10^{-3}$.}
		\label{fig:blast2}
	\end{center}
\end{figure}

\begin{figure}[!ht]
	\begin{center}
		\includegraphics[width=0.90\textwidth,clip=]{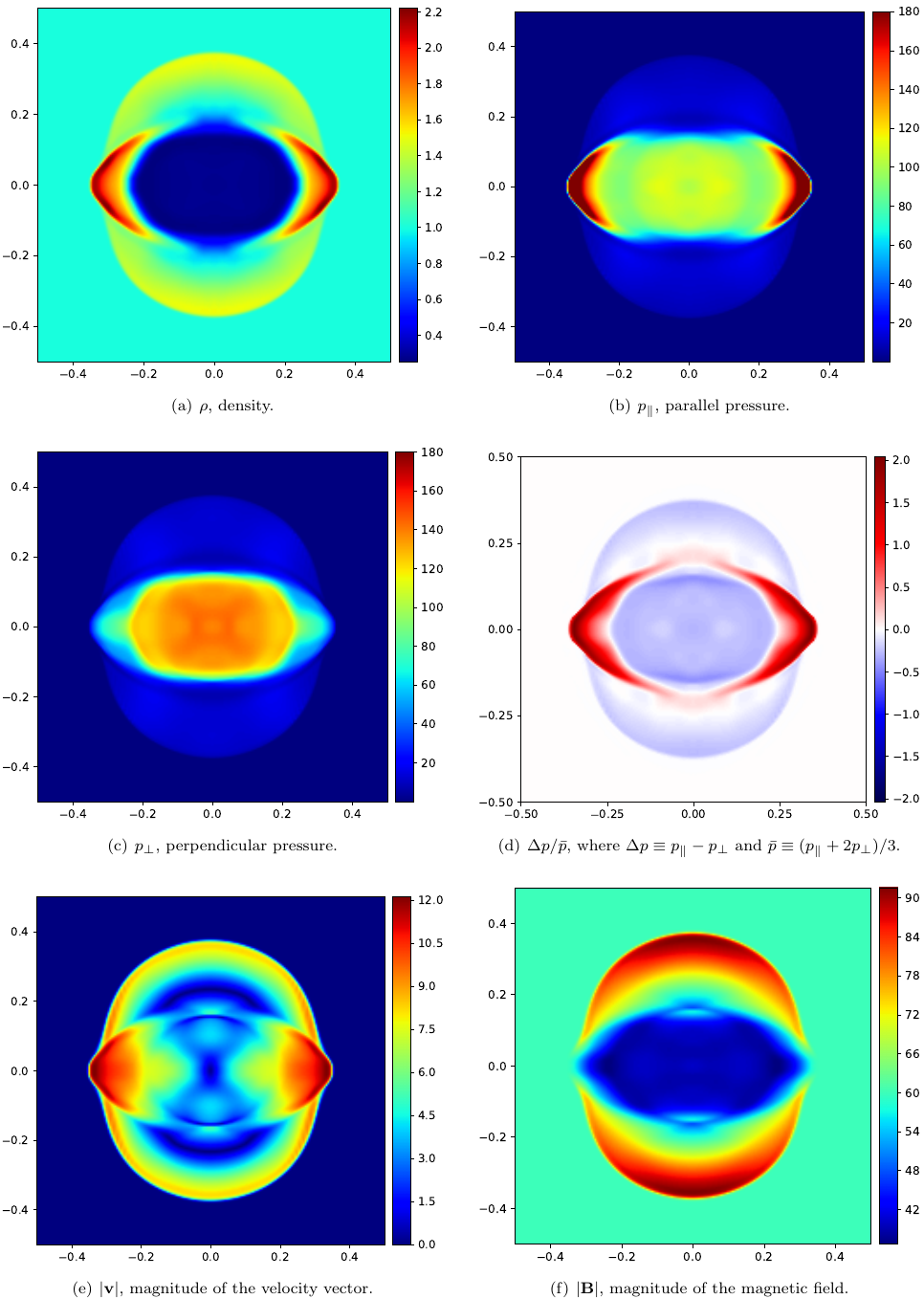}
		\caption{\nameref{subsec:blast}: The panels show the plots of $\rho$, $p_\parallel$, $p_\perp$, $\Delta p / \bar p$, $|\bf{v}|$ and $|\bf{B}|$ that correspond to $\tau _{phys} = 10^{-2}$.}
		\label{fig:blast3}
	\end{center}
\end{figure}

\begin{figure}[!ht]
	\begin{center}
        \includegraphics[width=0.90\textwidth,clip=]{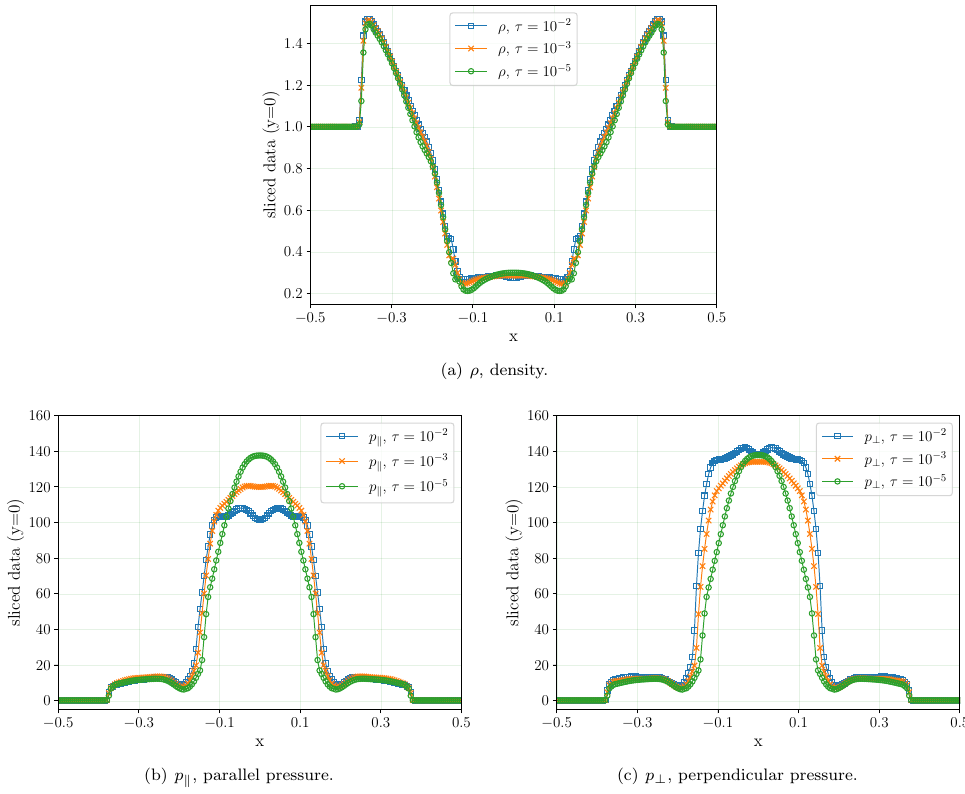}
		\caption{\nameref{subsec:blast}: Panels show the cut plots (along $y=0$) of $\rho$, $p_\parallel$, and $p_\perp$ profiles for various values of the relaxation time $(\tau_{phys}=10^{-2},10^{-3} \text{ and }10^{-5})$.}
		\label{fig:blast_slice}
	\end{center}
\end{figure}

\begin{figure}[!ht]
	\begin{center}
		\includegraphics[width=0.90\textwidth,clip=]{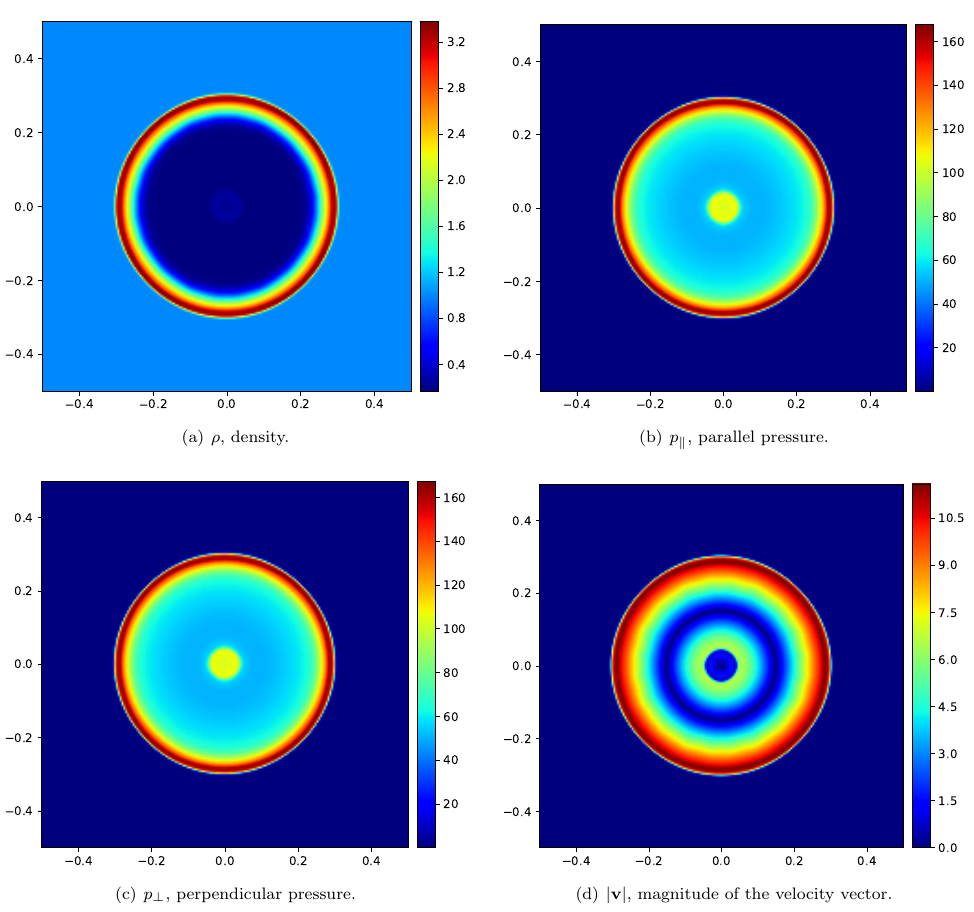}
		\caption{\nameref{subsec:blast}: The panels show the plots of $\rho$, $p_\parallel$, $p_\perp$ and $|\bf{v}|$  for the case where the initial magnetic field is set to zero (${\bf{B}}=0.0$).}
		\label{fig:blast_euler1}
	\end{center}
\end{figure}

We consider a CGL variation of the blast problem that was initially introduced for MHD in~\cite{balsara1999staggered}. The problem is configured on a unit square grid that consists of $200\times200$ zones and spans the domain $[-0.5, 0.5] \times [-0.5, 0.5]$. Initially, the density is uniformly set to unity across the domain. The parallel and perpendicular pressures are assumed to be equal and are uniformly set to 0.1 except within a central circle of radius 0.1, where they are elevated to 1000. Initially, the velocity is zero throughout. Additionally, a magnetic field with a magnitude of 60 is initialized along the $x-$direction. We simulate the problem until a final time of $t = 0.01$ using the second-order scheme and show results for various $\tau_{phys}$ in Figs.~\ref{fig:blast1}--\ref{fig:blast3}. \rev{For a better understanding of the shock-profiles, in Fig.~\ref{fig:blast_slice}, we plot horizontal cuts (along $y=0$) of the density, parallel pressure, and perpendicular pressure profiles for various values of $\tau_{phys}$.}

Fig.~\ref{fig:blast1} corresponds to $\tau _{phys} = 10^{-5}$ with the result that our physical relaxation time is substantially smaller than the timestep in this problem. We see, therefore, that the parallel and perpendicular pressures appear to be very isotropic. Fig.~\ref{fig:blast2} corresponds to $\tau _{phys} = 10^{-3}$ with the result that our physical relaxation time is much larger than the timestep in this problem. Fig.~\ref{fig:blast3} corresponds to $\tau _{phys} = 10^{-2}$ so that our physical relaxation time is comparable to the final simulation time in this problem. We see that with increasing $\tau _{phys}$ we have an increase in the pressure anisotropy, as expected.

As in the rotor example, the parallel and perpendicular pressures are identical at the start of the simulation, and the magnetic field is initially horizontal. As in the rotor problem, the pressure anisotropies have been generated spontaneously by the physics of the governing CGL equations. By looking at the top and bottom of the rotor in Figs.~\ref{fig:blast2}c and~\ref{fig:blast3}c, we see that the outward propagating wave is propagating into the horizontal field, causing a substantially greater increase in the perpendicular pressure compared to the parallel pressure. The blast has also compressed the inner part of the domain, with the result that the center of the blast also shows an enhanced perpendicular pressure. We can also focus on the horizontal ends of the blast wave in Figs.~\ref{fig:blast2}b and~\ref{fig:blast3}b, to observe that the parallel pressure is much enhanced there. \rev{The same behavior of the pressure profiles can be seen in the cut-plots given in Figs.~\ref{fig:blast_slice}b and~\ref{fig:blast_slice}c.}

In Sub-section~\ref{subsec:euler_lim}, we showed that the CGL equations, coupled with the concept of an "elastic fence" can smoothly produce an isotropic pressure in the limit where the magnetic field tends to (and reaches) zero in a code. Here, we numerically demonstrate this by setting the magnetic field to zero $(\mathbf{B}=0)$ in the above test case. We present the results in Fig.~\ref{fig:blast_euler1} and observe that our scheme yields identical parallel and perpendicular pressures, and all profiles exhibit a strong resemblance to the hydrodynamical limit. This test problem, numerically, illustrates that the CGL equations can indeed operate stably in the hydrodynamic limit.

\subsection{Orszag-Tang Problem for the CGL Equations} \label{subsec:ot}

\begin{figure}[!ht]
	\begin{center}
		\includegraphics[width=0.90\textwidth,clip=]{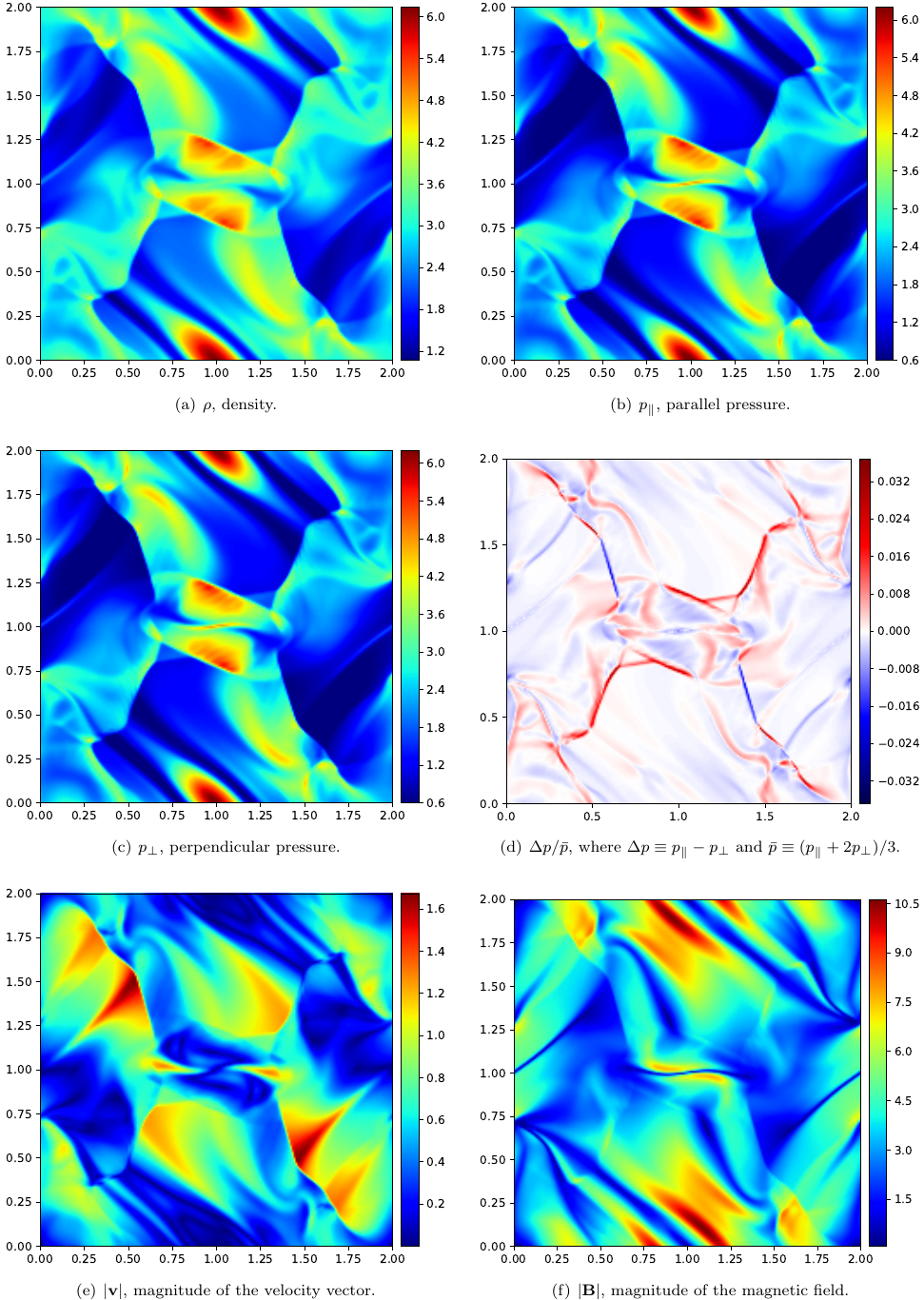}
		\caption{\nameref{subsec:ot}: The panels show the plots of $\rho$, $p_\parallel$, $p_\perp$, $\Delta p / \bar p$, $|\bf{v}|$ and $|\bf{B}|$ that correspond to $\tau _{phys} = 10^{-4}$.}
		\label{fig:ot1}
	\end{center}
\end{figure}

\begin{figure}[!ht]
	\begin{center}
		\includegraphics[width=0.90\textwidth,clip=]{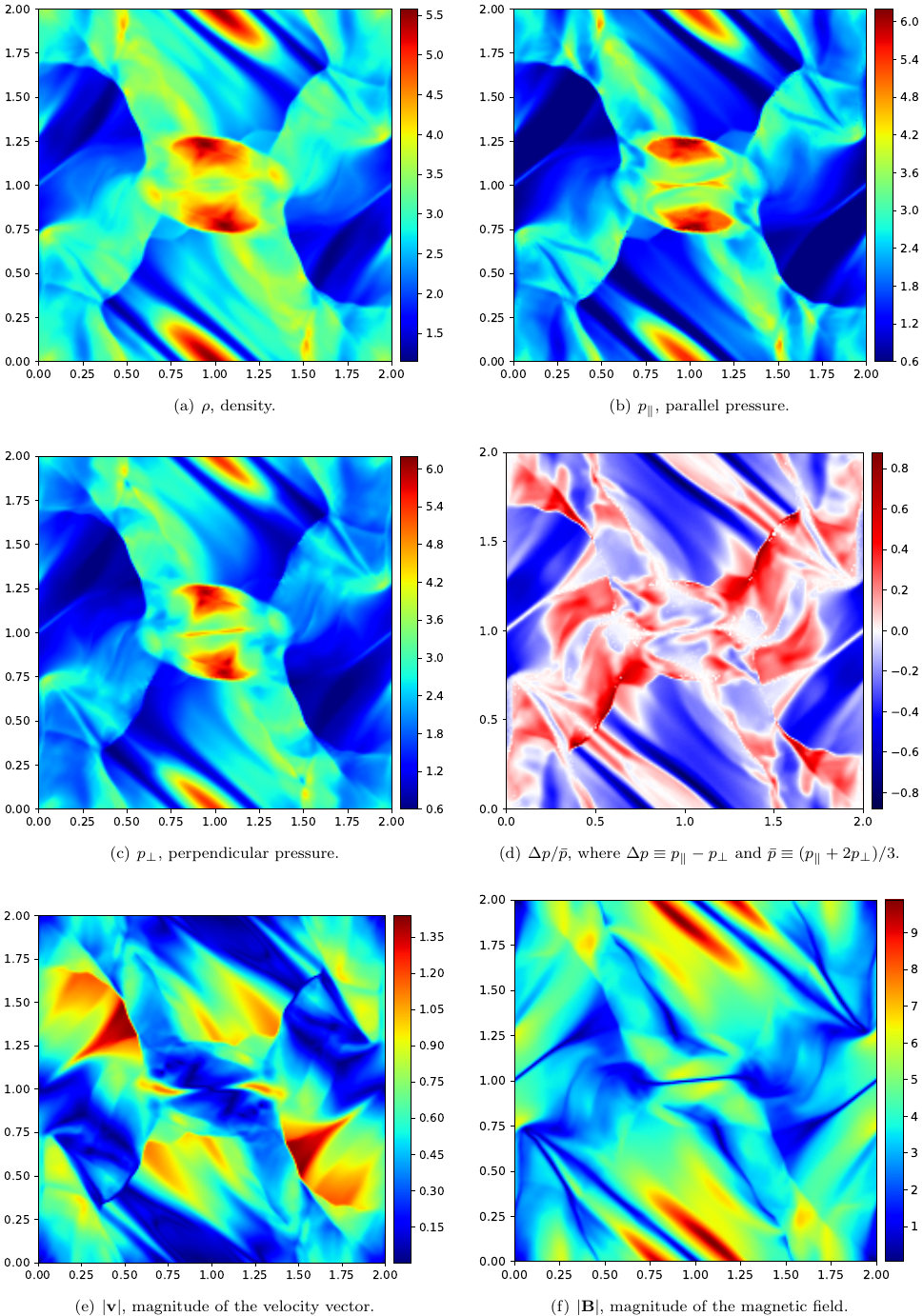}
		\caption{\nameref{subsec:ot}: The panels show the plots of $\rho$, $p_\parallel$, $p_\perp$, $\Delta p / \bar p$, $|\bf{v}|$ and $|\bf{B}|$ that correspond to $\tau _{phys} = 1$.}
		\label{fig:ot2}
	\end{center}
\end{figure}

The Orszag-Tang problem is usually taken to be a model for the transition to MHD turbulence. The problem was first introduced in~\cite{orszag1979small} (also see~\cite{balsara1998total}). The problem is initialized on a square domain $[0,2]\times[0,2]$ with the periodic boundaries. The density was initialized to unity all over the domain. The parallel and perpendicular pressures are assumed to be equal and are uniformly set to $5/3$. The velocity field is set up as follows:
$$\mathbf{v} = -\sin ( \pi y ) \hat{x} + \sin ( \pi x ) \hat{y}.$$
The magnetic field is initialized in a divergence-free manner using the following vector potential:
$$
A_z=-\dfrac{\sqrt{4 \pi}}{2 \pi}
\left(
\cos(2 \pi x) + 2\cos(\pi y)
\right).
$$
We run simulations until a final time of $t = 1$ on a grid of $200 \times 200$ zones. We use third-order scheme, and show results for various $\tau_{phys}$ in Figs.~\ref{fig:ot1}--\ref{fig:ot2}.

Fig.~\ref{fig:ot1} shows that the pressures are indeed isotropic when the relaxation time is very small ($\tau_{phys}=10^{-4}$). Fig.~\ref{fig:ot2} corresponds to a much larger relaxation time ($\tau_{phys}=1$). In that case we see the development of very anisotropic pressures. This gives us the perspective that in astrophysical and space plasmas where pressure anisotropies can develop, they will indeed contribute significantly to the structure of the turbulence. Since the pressure sets the temperature, and since the temperature regulates the emissivity of a plasma, this indicates that plasmas that are simulated with the CGL equations should, under the right conditions, show larger emissivity fluctuations compared to plasmas that are evolved with the MHD equations.

\subsection{Field Loop Advection Problem for the CGL Equations}\label{subsec:fieldloop}

\begin{figure}[!ht]
	\begin{center}
		\includegraphics[width=0.75\textwidth,clip=]{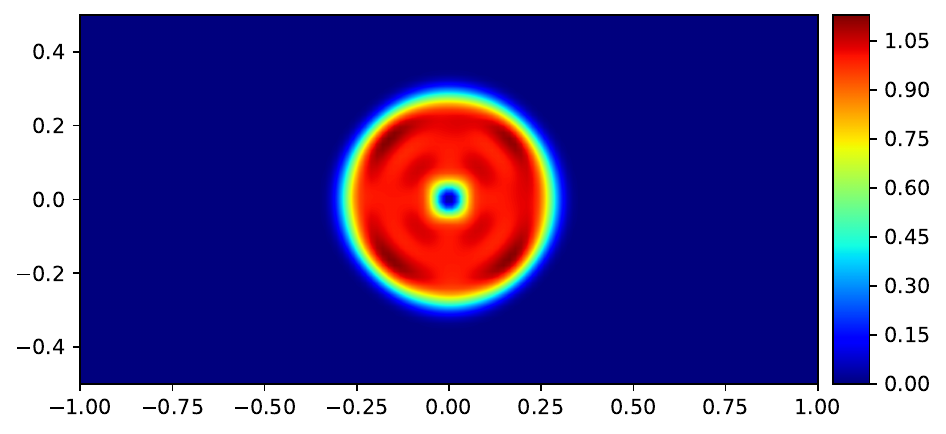}
		\caption{\nameref{subsec:fieldloop}: This figure shows the plot of $|\mathbf{B}|^2/(4\pi)$ that corresponds to $\tau _{phys} = 10^{-1}$.}
		\label{fig:fieldloop}
	\end{center}
\end{figure}

This problem is set up on a $128 \times 64$ zone domain that spans $[-1, 1] \times [-0.5, 0.5]$. The problem consists of advecting a two-dimensional loop of the magnetic field with a very low magnetic pressure compared to the gas pressure. The loop is advected along the diagonal of the computational domain with periodic boundaries applied in both directions. It was initially
described in~\cite{gardiner2005unsplit}, and a variation of this problem in the CGL framework has been done in~\cite{hirabayashi2016new}. We follow the set-up described in~\cite{hirabayashi2016new}, and do not repeat the description here. The conventional result examines the structure of the magnetic loop after it has completed one complete orbit around the domain. The problem was run to a stopping time of unity using fourth-order accurate scheme. 

Fig.~\ref{fig:fieldloop} shows the magnitudes of the magnetic field for the field loop advection problem after it has executed one complete orbit along the diagonal of the computational domain. We use an intermediate relaxation time $\tau_{phys}=10^{-1}$ for this problem. We see that there is virtually no diffusion of the loop's boundaries and no oscillations in the magnetic pressure within the loop's interior. We, therefore, conclude that the method presented here has adequate dissipation behavior for the advection of magnetic fields.

\subsection{Shock interaction with a magnetosphere}\label{subsec:shockdipole}

\begin{figure}[!ht]
	\begin{center}
		\includegraphics[width=0.85\textwidth,clip=]{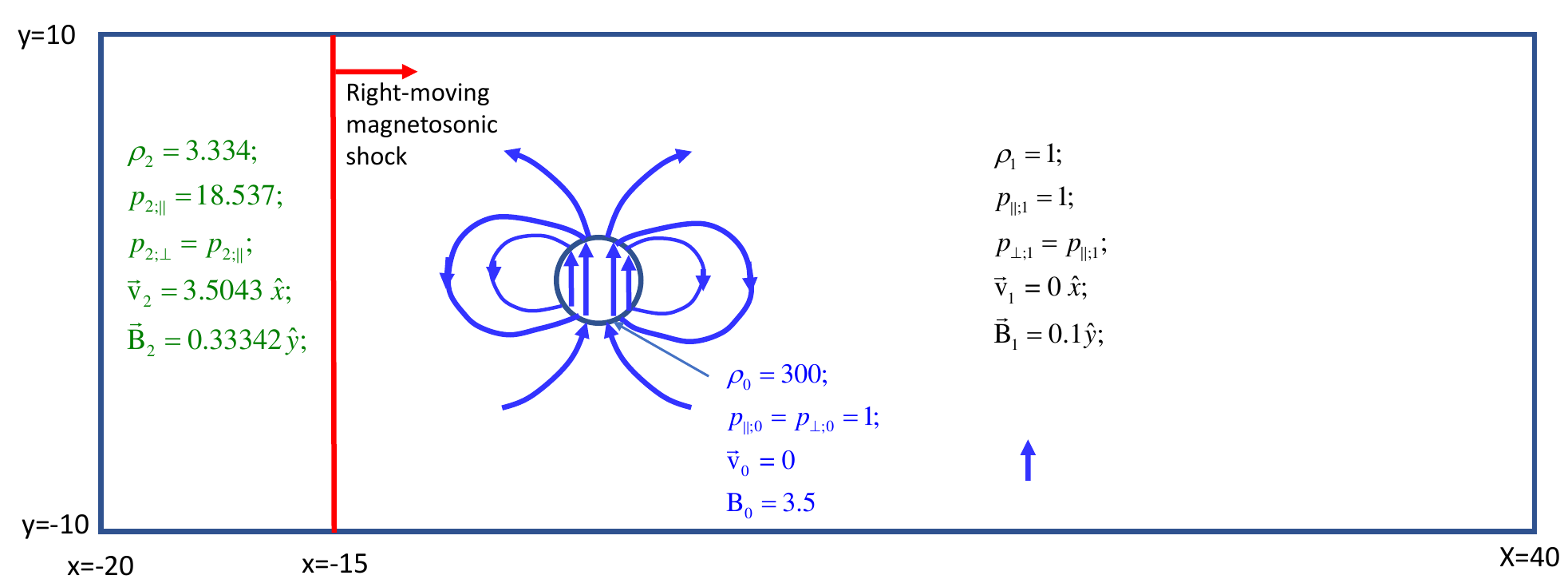}
		\caption{\nameref{subsec:shockdipole}: The schematic figure illustrates the setup of a two-dimensional test problem that very roughly imitates a shock interacting with a magnetosphere. The domain limits are shown in the figure. A cylinder of unit radius that is threaded on the inside with a magnetic field with magnitude $B_0$ is also shown. The parameters that are initially used for the cylinder and the field are shown in the figure. The exterior field that threads the cylinder is intended to look like a dipolar field in 2D. A uniform magnetic field with $B_1 =  0.1$ threads the region that makes up the ambient medium around the cylinder. The post-shock variables are also shown in the figure. The right-going fast magnetosonic shock eventually overruns the magnetized cylinder.}
		\label{fig:shockdipole_data}
	\end{center}
\end{figure}

In this sub-section, we consider a two-dimensional test problem that very roughly imitates a shock interacting with a magnetosphere. The two-dimensional computational domain spans $\left( x,y \right)\in \left[ -20,40 \right]\times \left[ -10,10 \right]$. The set-up data is schematically shown in Fig.~\ref{fig:shockdipole_data}. A cylinder with a unit radius is set up at the origin. The density inside the cylinder is ${{\rho }_{0}}=300$ with pressures within the cylinder initialized as ${{p}_{\parallel 0}}={{p}_{\bot 0}}=1$. A constant magnetic field is initialized within the cylinder in the y-direction with ${{B}_{0}}=3.5$. This cylinder is embedded in a much lighter ambient fluid with density ${{\rho }_{1}}=1$ and pressures ${{p}_{\parallel 1}}={{p}_{\bot 1}}=1$. The fluid that makes up the cylinder and its ambient are initially not moving. External to the cylinder, we set up a magnetic structure that looks like a dipole in 2D with magnetic field given by
\begin{equation}
    \mathbf{B}=(B_x,B_y,B_z)^\top=
    \left(
    B_0 \dfrac{2xy}{(x^2+y^2)^2},
    B_0 \dfrac{(y^2-x^2)}{(x^2+y^2)^2},
    0
    \right)^\top
\end{equation}
The corresponding vector potential, which is useful for initializing the problem, is given by
\begin{align}
    {\vec{A}}_{0}(x,y) = 
    \begin{cases}
      -{{B}_{0}}x  \hat{z} & \text{for} \  {{x}^{2}}+{{y}^{2}}\le 1\\
      -\dfrac{{{B}_{0}}x}{{{x}^{2}}+{{y}^{2}}}  \hat{z} & \text{for} \  {{x}^{2}}+{{y}^{2}} > 1
    \end{cases}
\end{align}
The magnetic field that is set up external to the cylinder is designed to dovetail with the magnetic field within the cylinder. Because we intend to set up a right-going magnetosonic shock that eventually overruns the cylinder, we also add to the cylinder’s internal field, as well as to its ambient magnetic field, an additional constant magnetic field ${{\vec{B}}_{1}}=0.1\hat{y}$. This ensures that the shocked gas will have an even larger y-component of the magnetic field. This right-going fast magnetosonic shock is initialized along the line $x=-15$. The shocked gas has density  ${{\rho }_{2}}=3.334$, pressures ${{p}_{\parallel 2}}={{p}_{\bot 2}}=18.537$, velocity ${{{\vec{v}}}_{2}}=3.5043\hat{x}$ and magnetic field ${{\vec{B}}_{2}}=0.33342\hat{y}$. The boundary conditions are periodic in the y-direction; inflow at the left boundary and outflow at the right boundary. The problem was run to a simulation time of 10 units, by which time the shock has overrun the dense cloud and is close to the right boundary. A mesh of $1200\times 400$ zones was used.

\begin{figure}[!ht]
	\begin{center}
		\includegraphics[width=0.95\textwidth,clip=]{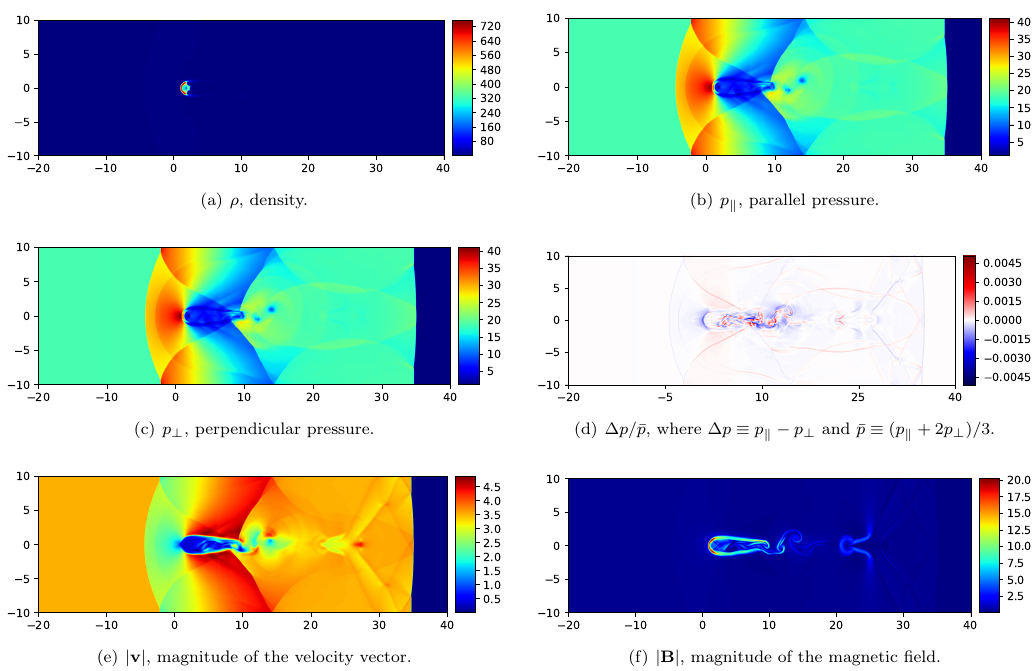}
		\caption{\nameref{subsec:shockdipole}: Panel shows the plots of $\rho$, $p_\parallel$, $p_\perp$, $\Delta p / \bar p$, $|\bf{v}|$ and $|\bf{B}|$ that correspond to $\tau _{phys} = 10^{-4}$.}
		\label{fig:dipoleshock1}
	\end{center}
\end{figure}

\begin{figure}[!ht]
	\begin{center}
		\includegraphics[width=0.95\textwidth,clip=]{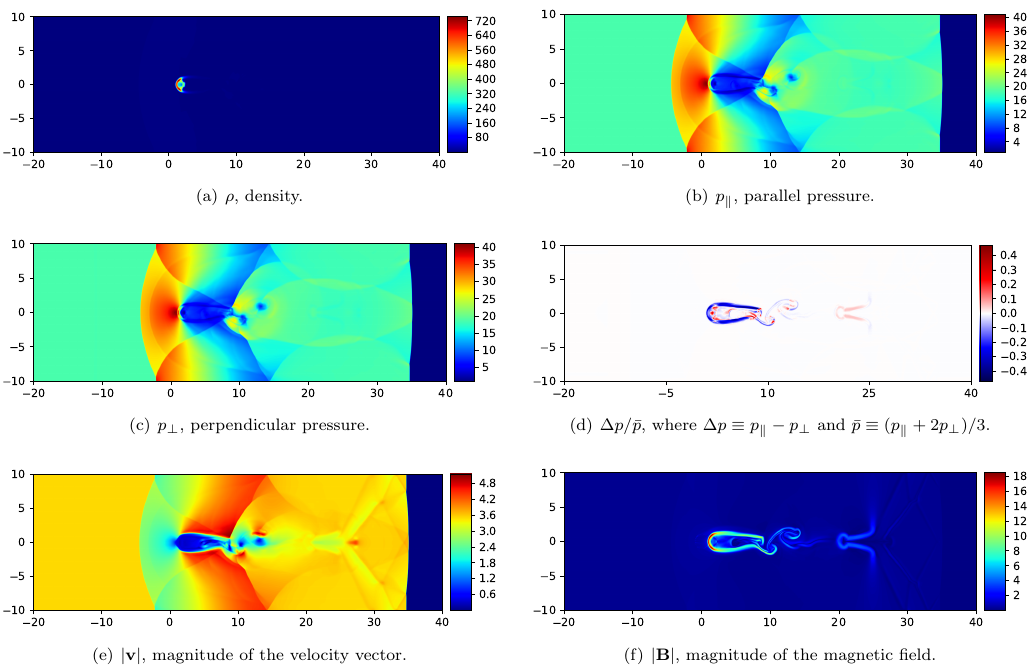}
	    \caption{\nameref{subsec:shockdipole}: Panel shows the plots of $\rho$, $p_\parallel$, $p_\perp$, $\Delta p / \bar p$, $|\bf{v}|$ and $|\bf{B}|$ that correspond to $\tau _{phys} = 1$.}
		\label{fig:dipoleshock2}
	\end{center}
\end{figure}

Figs.~\ref{fig:dipoleshock1} and~\ref{fig:dipoleshock2} show two versions of this problem with ${{\tau }_{phys}}={{10}^{-4}}$ (i.e. the MHD limit) and ${{\tau }_{phys}}=1$ (i.e. the CGL limit). Several common features are shared between the two simulations, as expected, but we can also pick out prominent differences. Compare Figs.~\ref{fig:dipoleshock1}d and~\ref{fig:dipoleshock2}d. We see that there is substantially more pressure anisotropy in Fig.~\ref{fig:dipoleshock2}d as compared to Fig.~\ref{fig:dipoleshock1}d. From Fig.~\ref{fig:dipoleshock2}d we see that the part of the bow shock that faces into the incoming flow displays a ${{p}_{\bot }}$ that is substantially larger than ${{p}_{\parallel }}$, which is consistent with observations and also physical expectations. In the magnetotail region we see parts of the flow with ${{p}_{\bot }}>{{p}_{\parallel }}$ as well as parts of the flow with ${{p}_{\bot }}<{{p}_{\parallel }}$ indicating that the pressure fluctuations in the CGL limit could strongly contribute to the turbulence in the magnetotail. This is just a 2D test problem, so it is not expected to produce fully-developed turbulence. Even so, comparing the velocity plots in Figs.~\ref{fig:dipoleshock1}e and~\ref{fig:dipoleshock2}e, we see that the velocity structure in the magnetotail region shows substantially enhanced turbulence in Fig.~\ref{fig:dipoleshock2}e compared to Fig.~\ref{fig:dipoleshock1}e. Despite being a 2D simulation, we see several very significant trends that we will follow up in 3D and with more realistic conditions.

\section{Conclusions} \label{sec:conc}
MHD is built on assuming that there is a single isotropic pressure. Several lines of observational evidence, stemming from the solar heliosphere and the magnetospheres of planets in our solar system, have demonstrated that the plasma pressures can often become anisotropic. Astrophysical systems, like accretion disks and the intercluster medium, also bring out the need for modeling pressure anisotropies in astrophysical plasmas. The CGL system is the simplest example of a hyperbolic PDE system that allows plasma pressures to become anisotropic. 

We have made a detailed study of the CGL system and the challenges it poses. At a numerical level:- 1) the system is in non-conservation form, 2) it has stiff source terms, 3) it is prone to a loss of hyperbolicity. Furthermore, since different physical instabilities come into play when ${{p}_{\parallel }}>{{p}_{\perp }}$, and when ${{p}_{\parallel }}<{{p}_{\perp }}$, it is imperative that the operation of the stiff source terms also preserve these inequalities. The numerical challenges cannot be solved without detailed consideration of the physics of the instabilities that dominate a collisionless plasma. For that reason, we realize that when the pressure anisotropy becomes too large, and with ${{p}_{\parallel }}>{{p}_{\perp }}$, the firehose instability will dominate. On the other hand, when the pressure anisotropy becomes too large, and with ${{p}_{\parallel }}<{{p}_{\perp }}$, the mirror instability will dominate. These instabilities can operate on the microscales of a large-scale computation and drive the system towards isotropy. We realize, therefore, that physics and numerics need to work hand-in-glove in order to make the CGL equations numerically tractable. To that end, we first identify the limits to the pressure anisotropy past which the firehose and mirror instabilities act. To include that physics in the numerics, we invent the idea of an “elastic fence”. It ensures that if the system approaches the “hard fence” associated with these instabilities, it is gently pushed away from hitting these limits. Indeed, in order to function reliably, a numerical code must stay away from these limits because if it did not, hyperbolicity would be violated and the code would crash immediately because it would not be able to evaluate the physical eigenvalues. As inability to evaluate eigenvalues translates into an inability to evaluate the next timestep in any code for the CGL equations. We see, therefore, that any numerical code for the CGL equations should stay safely clear of the “hard fence” by resorting to the “elastic fence” that we propose.

To accommodate this blend of physics-based and numerically-driven requirements, we invent a bounds-preserving time-integration strategy that is stable even in the presence of ultra-stiff source terms. The bounds-preserving aspect of the scheme ensures that the sign of $\left( {{p}_{\parallel }}-{{p}_{\perp }} \right)$ is preserved by an application of the stiff source terms. (Note, however, that the non-conservative products in the equation for ${\partial \left( {{p}_{\parallel }}-{{p}_{\perp }} \right)}/{\partial t}\;$ can indeed drive the system away from isotropy and we design accurate methods that allow that to happen if the physical problem has to go in that direction.) We also recast the system in fluctuation form so that the system retains conservation when that is warranted while accommodating the presence of non-conservative products. The “elastic fence” is, therefore, a way of emulating in a numerical code the same thing that nature chooses to do via the operation of the firehose and mirror instabilities. 

Several practical ideas are presented for conserving thermal energy while keeping the pressure anisotropy within physically acceptable bounds. In practice, it means that the CGL equations by themselves do not have the requisite “knowledge” to stay within the domain of physical realizability, and our innovations provide the extra “guidance” to ensure that this is achieved. Details on the bounds-preserving timestepping, the fluctuation form, the Riemann solvers that support this fluctuation form, and the design of the “elastic fence” are presented in this work. Section~\ref{sec:stepbystep} brings all these innovations together by explaining how the different steps are integrated into numerical code.
 
A completely non-trivial test problem is constructed in Section~\ref{sec:accuracytests} to show that the proposed numerical methods for the CGL equations meet their design accuracy. We demonstrate that our numerical code can reach its designed second, third, and fourth-order accuracy. We also present several stringent numerical examples. The test problems illustrate that when the physical relaxation time is small, pressure isotropy is indeed very closely achieved by the CGL equations – in other words, the CGL equations can access the MHD limit if our proposed methods are used. However, when the physical relaxation time is large, the CGL equations show us the different types of physics that arise when pressure isotropy is violated. It has also been claimed in the literature that the CGL equations cannot reclaim the hydrodynamic limit because $\mathbf{b}={\mathbf{B}}/{\left| \mathbf{B} \right|}\;$ becomes undefined. But we show that this claim is specious in light of the fact that the physics of the “elastic fence” restores pressure isotropy when the plasma beta becomes extremely large. As a result, we demonstrate, perhaps for the first time, that a CGL code will stably operate in the hydrodynamical limit. We also include a test problem that mimics the interaction of a shock with a magnetospheric environment in 2D.

\section{Acknowledgments}
This work was supported by NSF grant 2009871 and by NASA grant 80NSSC20K0786.

\bibliography{ms}{}
\bibliographystyle{aasjournal}

\end{document}